\documentclass[11pt,a4paper]{article}

\usepackage{amsmath,amsfonts,latexsym,amssymb}
\usepackage{verbatim,amsthm,graphics}
\usepackage{mathrsfs}

\def\ben{\begin{equation}}
\def\een{\end{equation}}
\def\bena{\begin{eqnarray}}
\def\eena{\end{eqnarray}}
\def\non{\nonumber}

\def\mS{{\mathbb S}}
\def\mV{{\mathbb V}}

\begin{document}
\title{Massive vector field perturbations on extremal and near-extremal static black holes}

\author{
Kodai Ueda$^{1}$\thanks{\tt 1733310105r@kindai.ac.jp} and 
Akihiro Ishibashi$^{1,2}$\thanks{\tt akihiro@phys.kindai.ac.jp} \\ 
{} \\ 
{\it ${}^{1}$Department of Physics, and ${}^{2}$Research Institute for Science and Technology}
{}\\
{\it Kindai University, } \\
{\it Higashi-Osaka, Osaka, 577-8502, Japan} 
}

\date{\today}

\maketitle

\begin{abstract}

We discuss a new perturbation method to study the dynamics of massive vector fields on extremal and near-extremal static black hole spacetimes. 
We start with, as our background, a rather generic class of warped product metrics that 
consist of an $m$-dimensional spacetime and an $n$-dimensional Einstein space, and with respect to the behavior on the latter space, we 
classify the components of massive vector fields into the vector(axial)- and scalar(polar)-type components. On this generic background spacetime, we show that for the vector-type components, the Proca equation reduces to a single homogeneous 
master equation, whereas the scalar-type components remain coupled. 
Then, focusing on the case of extremal and near-extremal static black holes in four-dimensions, we consider 
the near-horizon expansion of both the background geometry and massive vector field by a scaling parameter $\lambda$ 
with the leading-order geometry of $\lambda \rightarrow 0$ 
being the so called near-horizon geometry. 
We show that on the near-horizon geometry, thanks to its enhanced symmetry, the Proca equation for the scalar-type components also 
reduces to a set of two mutually decoupled homogeneous wave equations for two scalar variables, plus a coupled equation through which 
the remaining third variable is determined from one of the first two. 
Therefore, together with the vector-type master equation for a single variable, we obtain the set of three decoupled master wave equations, 
each of which governs each of the three independent dynamical degrees of freedom of the massive vector field in four-dimensions. 
We further expand the geometry and massive vector field with respect to $\lambda$ and 
show that at each order of $\lambda$, the Proca equation for the scalar-type components can reduce to a set of two mutually  
decoupled inhomogeneous wave equations whose source terms consist only of the lower-order variables, plus 
one coupled equation that determines the remaining third variable. 
Therefore, if one solves the master equations on the leading-order near-horizon geometry, 
then in principle one can successively solve the Proca equation at any order.

\end{abstract}

\section{Introduction}
\label{sect:Intro}

Exploring the dark sector of our universe is one of the major, most challenging subjects 
in recent cosmophysics. In particular, identifying the dark sector of matter is of great importance in the context of astrophysics and high-energy particle physics. 
One of the most popular candidates for dark matter are weakly coupled ultra-light bosonic fields, predicted to generally arise 
in string-theory-inspired scenarios~\cite{Arvanitaki:2009fg,Acharya:2015zfk,Goodsell:2009xc}. One way to search for such 
ultra-light bosonic fields in astrophysical context is to seek for the so called {\em superradiant instability} of rotating black holes, which produces gravitational radiation and eventually spins down the black hole~\cite{Arvanitaki:2010sy}.  
A superradiant scattering around a rotating black hole can in general occur when the frequency $\omega$ 
of an impinging wave of some bosonic fields satisfies the condition 
$0<\omega < m\Omega_H$, where $m$ denotes the azimuthal number of the impinging wave 
and $\Omega_H$ the horizon angular velocity of the black hole. 
When, in addition, there is some mechanism that reflects the superradiantly scattered, amplified wave back into the ergoregion of the black hole, the superradiant amplification can take place repeatedly and exhibit an instability. 
The reflection mechanism can be played by, for example, a ``mirror'' set 
by hand~\cite{Press:1972zz,Cardoso:2004nk,Herdeiro:DR2013,Hod:2013a,Degollado:Herdeiro:2014,LiZhao:2015}, 
or the spacetime curvature produced 
by a negative cosmological constant~\cite{Hawking:Reall:2000,Cardoso:2004hs,Cardoso:2006wa,Uchikata:2009zz,Kodama:2009rq,Cardoso:Dias:Hartnett:Lehner:Santos:2014,Green:Hollands:Ishibashi:Wald:2016}. 
A superradiant instability can also be realized in more realistic astrophysical circumstances if the impinging bosonic fields posses 
masses~\cite{Damour:1976kh,Zouros:1979iw,Detweiler:1980uk,Dolan:2007mj,Rosa:2009ei,Hod:2011,Pani:2012vp,Pani:2012bp,Witek:2012tr,Brito:CP:2013,Yoshino:Kodama:2015,Hod:2016,Ishibashi:Pani:Gualtieri:Cardoso:2015}, 
and can be most efficient when the corresponding compton wavelength of the massive fields 
is comparable to the black hole radius. This is the case in which, for example, the masses are less than $10^{-10}$ eV, for the stellar mass black hole case. The existence of such ultra-light bosons has been suggested by, for example, the string axiverse 
scenario~\cite{Arvanitaki:2009fg}, and there have been a number of attempts to derive bounds on the masses of such ultra-light 
bosons by exploiting the recent developments of precision black hole physics, see, e.g. Refs.~\cite{Arvanitaki:2010sy,Pani:2012vp,Pani:2012bp,Cardoso:2018tly}. 

\medskip 

It is also worth mentioning that the presence of massive bosonic fields can affect the environment of stationary black holes 
by endowing with ``hair.'' In fact, for a certain configuration of massive complex scalar and vector fields, 
the possibility of hairy rotating black holes has been pointed out~\cite{Herdeiro:Radu2014,Herdeiro:Radu2017}. 
Instability of such hairy black holes has also been discussed~\cite{Ganchev:Santos:2017,Degollado:Herdeiro:Radu:18}. 

\medskip

To explore the possibility of superradiant instability in our universe and other possible roles of 
massive bosonic fields in astrophysics and fundamental physics, 
it is of considerable importance to understand precisely how such massive fields propagate in black hole spacetimes. 
There have been a number of relevant work along this line by using 
both analytical and numerical methods, and the behavior of the massive scalar fields has now been well-understood. 
As for massive vector and tensor fields, however, the analyses become much more involved.  
For example, in the massive vector field case, 
having three independent physical degrees of freedom due to the lack of gauge freedom, the equation for massive vector fields or Proca equation does not appear to be immediately separable in Kerr black hole background, 
let alone reducing to master equations, i.e.,  a set of decoupled second-order wave equations. 
This situation should be compared with the case of {\em massless} vector, i.e., Maxwell, field, for which Maxwell-equations are separable and further reduce to the Teukolsky's master equation. See also~\cite{Lunin:2017} for the separability of Maxwell's equations with a new ansatz for the 
gauge field as well as the analysis in higher dimensional rotating black hole backgrounds. 
Such a complexity of massive vector fields remains to be the case even for non-rotating, static black hole background. For example, in Schwarzschild spacetime, although the radial and angular parts are immediately separable thanks to the spherical symmetry, 
the Proca equation still does not appear to reduce to a set of decoupled master equations~\cite{Konoplya:2006,Konoplya:Zhidenko:Molina:2007,Rosa:Dolan:2012}.  
For these reasons, in order to study the behavior of massive vector fields, one has to appeal to a combination 
of some approximation and numerical method, or full numerical computation~\cite{Witek:2012tr,Zilhao:Witek:Cardoso:2015,East:Pretorius:2017}. 
As for some approximation method, for example, by extending Kojima's pioneering work~\cite{Kojima:92,Kojima:apj:93,Kojima:ptp:93}, 
the slow-rotation approximation for linear perturbations of scalar and massive vector fields on slowly rotating black holes has been formulated in~\cite{Pani:2012vp,Pani:2012bp}.  

\medskip 
  
The superradiant instability is regulated by the dimensionless parameter $\mu M$ (in unit $G=c=1$) 
with $\mu$ and $M$ being the masses, respectively, of the bosonic field and black hole, 
and is expected to be strongest e.g., when $\mu M \sim 1$ 
for maximally spinning extremal black holes. 
There have already been several observations that indicate the existence of highly spinning, nearly extremal black holes 
in our universe, see e.g., Refs.~\cite{McClintock:etal:2006,Middleton:16}.   
It is therefore of considerable interest to develop some formulas which can be applied for rapidly spinning, 
near-extremal or maximally rotating extremal black holes. In particular, so far little has been done for {\em analytically} studying 
the dynamics of massive vector fields in extremal and near-extremal black holes.

\medskip 

It is well-known that an extremal black hole admits what is called the {\em near-horizon geometry}, which is obtained by taking a certain scaling limit around the horizon neighborhood, 
and which admits an enhanced isometry higher than that of the original extremal black hole geometry~\cite{Bardeen:Horowitz:1999,KLR07}. 
For the maximally rotating extremal Kerr black hole, the near-horizon geometry--also called NHEK--has the enhanced symmetry ${\rm SL}(2,R)\times {\rm U}(1)$, which has been exploited to formulate a type of the gauge-gravity duality, called Kerr/CFT correspondence~\cite{Guica:Hartman:Song:Strominger:2009}. 
Further, the enhanced symmetry of the near-horizon geometry has recently been used to analytically compute radiation emissions 
from the near-horizon region of extremal Kerr black  holes~\cite{Porfyriadis:Strominer:2014,Porfyriadis:Shi:Strominger:2017}. 
For further study of the near-horizon geometries and their classification, see e.g., Refs.~\cite{Figueras:Kunduri:Lucietti:Rangamani:2008,Kunduri:Lucietti:2009,Hollands:Ishibashi:2010,Kunduri:Lucietti:2013} 
and references therein.

\medskip 
Apart from astrophysics context, massive bosonic fields around extremal and near-extremal 
black holes have received attention also in some theoretical contexts. 
This is the case also for the static extremal black holes. 
For instance, a superradiant scattering can also occur in non-rotating, static black holes 
if one considers a charged scalar field coupled to the background gauge field in a 
charged black hole. In this case, the role of $m\Omega_H$ for the rotating case is played by $ q \Phi_H$, where $q$ denotes the charge 
of the field and $\Phi_H$ the electric potential at the horizon. 
For example, the behavior of charged scalar fields in extremal Reissner-Nordstrom black holes has been studied by 
using the near-horizon geometry~\cite{Zimmerman:2017}. 
Another interesting phenomena is the condensation of massive scalar fields around the horizon of 
near-extremal black holes. 
The near-horizon geometry of an extremal black hole includes in part a two-dimensional anti-de Sitter (AdS$_2$) spacetime.  
Then, if the mass of massive field violates the so called Breitenlohner-Freedman bound \cite{BF1,BF2} of the near-horizon AdS$_2$ spacetime, 
massive scalar field condensates and triggers an instability, which may be interpreted as a phase transition in the dual boundary field theory.  
Such an instability due to the scalar field condensation is known to occur for 
near-extremal Reissner-Nordstrom-anti-de Sitter black hole, and has an interesting application to holographic superconductors~\cite{Gubser2008,Hartnoll:Herzog:Horowitz:2008}. 
For further study of stability of extremal and near-extremal black holes, see, e.g., Refs.~\cite{Dias:Monteiro:Reall:Santos:2010,Aretakis:2011,Durkee:Reall:11,Hollands:Ishibashi:15} and references therein. 

\medskip 

The purpose of this paper is to develop a novel perturbation method that can apply for studying the dynamics of massive vector fields 
in extremal and near-extremal black hole spacetimes.  
In an extremal and near-extremal black hole geometry, one can introduce a constant scaling parameter, say $\lambda$, which effectively plays a roll of zooming up the neighborhood of the horizon, and taking the limit $\lambda \rightarrow 0$ provides the near-horizon geometry. 
Our strategy is as follows. We first view $\lambda$ as a small parameter to expand the background extremal (or near-extremal) geometry around the near-horizon geometry. 
Next, on this expanded geometry we consider massive vector 
fields as perturbation with small amplitude parametrized again by $\lambda$. Then, we examine the Proca 
equations on the expanded geometry at each order of $\lambda$. 
Our approach may be viewed as a two-parameter perturbation in which 
both the amplitude of the metric and that of the massive vector field are small and simultaneously parametrized by $\lambda$. 
It should be noted that in our formulation, we assume that the geometry is already fixed in full order of $\lambda$ as a solution of the Einstein equations (i.e., not required to be solved at each order of $\lambda$), 
while Proca perturbations are our dynamical variables to be solved at each order. 

\medskip

In this paper, as a first step toward formulating our ideas stated above, we restrict our attention to static (near-) extremal black holes. 
One advantage in the static case is that the radial and angular parts of the Proca equations can immediately be 
separated as we shall demonstrate explicitly below. 
We first show, after the separation of variables, that the Proca equation for massive vector field 
can be reduced to a set of decoupled master equations (for three-independent dynamical degrees of freedom 
in four-dimensions) at leading order of $\lambda$, i.e., on the near-horizon geometry. 
Next, we expand both the geometry and massive vector field with respect to the small parameter $\lambda$, 
and obtain the formulas for higher order perturbation equations. We show that at each order 
of $\lambda$, one can obtain a set of mutually decoupled wave equations, each of which governs 
each independent dynamical degree of freedom, and each of which has a source term consisting only 
of the lower-order variables. Thus, in principle, starting from solving the leading order decoupled 
master equations, one can iteratively solve any order of massive 
vector perturbations. As a concrete example, we present the relevant formulas in the (near-) extremal 
Reissner-Nordstrom background spacetime. 
It is worth commenting that although the focus of this paper is on the static case, as the separation of variables 
can be easily performed, it was recently shown that even for the rotating case, equations for massive vector fields 
are separable~\cite{Frolov:Krtous:Kubiznak:Santos:2018}. We therefore expect that our method developed 
in this paper may be generalized to the maximally rotating black hole case. 

\medskip 

In the next section, we describe our background geometry and the Proca equation, thereby 
establishing our notation. Our background metric takes the warped product form with 
an $m$-dimensional spacetime and an $n$-dimensional Einstein space. 
We classify the components of massive vector 
field into two types: the divergence-free {\em vector(axial)}-type part and 
{\em scalar(polar)}-type part with respect to the behavior on the Einstein space. By doing so, we can 
deal with the vector- and scalar-type parts separately.  
Then we introduce scalar and vector harmonics on the Einstein space and make the separation 
of variables into the radial and angular parts. Then, we reduce the Proca equation to a set of 
wave equations on the $m$-dimensional spacetime. At this stage, for the vector-type component, 
we obtain a single master equation, whereas for the scalar-type components, 
the equations are still coupled. We also discuss the massless case, and present a master equation 
for the scalar-type components of Maxwell field. 
In section~\ref{sect:Extremal}, we describe extremal and near-extremal black 
hole spacetimes, introduce the extremality parameter as well as the near-horizon scaling parameter, 
of which zero-limit corresponds to the near-horizon geometry. 
In section~\ref{sect:Expanding}, we formulate our perturbation method of expanding both the field variables and 
the background geometry with respect to the scaling parameter, and derive our main formulas at each 
order of perturbation.  
We first do so for the standard four-dimensional (near-) extremal 
Reissner-Nordstrom background, and then for a general warped product type background. 
We also explicitly give the general solutions to the leading-order wave equations for both vector- and 
scalar-type components. We show that the same structure 
for massive vector field perturbations also holds in the more generic 
extremal and near-extremal static black hole background. 
Section~\ref{sect:Summary} is devoted to summary and discussion. 
For completeness, we also apply our perturbation method to charged massive scalar fields 
in (near-)extremal Reissner-Nordstrom black hole background, and derive the relevant formulas 
in Appendix. 

\section{Background geometry and Proca equations}
\label{sect:Background}

Although our main concern is the dynamics of massive vector fields 
in four-dimensional black hole spacetime, taking 
into consideration the possibility of a wide variety of applications in fundamental physics,  
we shall present the relevant formulas in a rather generic setup. 
We first describe our warped product type background geometry 
and next discuss how to classify vector fields 
on our background. This part largely follows Refs.~\cite{Kodama:Ishibashi:2003,Ishibashi:Kodama:2003}.  
We then write down the Proca equations explicitly in our background spacetime.

\subsection{General warped product type geometry} 
Let us consider $D=(m+n)$-dimensional spacetime whose manifold structure is given locally 
as a warped product 
${\cal M} = {\cal N}^m \times {\cal K}^n$. 
We distinguish tensors living in each manifold $M,{\cal N}^m,{\cal K}^n$ 
by using greek indices for tensors on ${\cal M}$,  latin indices in the range $a,b,c,\dots$ on ${\cal N}^m$, latin indices in the range $i,j,k\dots$ on ${\cal K}^n$. Accordingly we introduce local coordinates 
on ${\cal M}$ as $x^\mu=(y^a,z^i)$ so that the metric takes the following form: 
\ben
ds^2 =g_{\mu \nu}dx^\mu dx^\nu = {}^{m}g_{ab}(y)dy^ady^b + R^2(y) \gamma_{ij}(z)dz^i dz^j  \,, 
\label{def:background}
\een 
where ${}^{m}g_{ab}(y)$ and $\gamma_{ij}(z)$ denotes, respectively, the Lorentzian metric on ${\cal N}^m$ and  
Riemannian metric on ${\cal K}^n$. We further assume that $({\cal K}^n,\gamma_{ij})$ is the $n$-dimensional Einstein space, so that its Ricci curvature satisfies $\hat{R}_{ij} = K(n-1)\gamma_{ij}$, with $K=0, \pm 1$ denoting the sectional curvature of ${\cal K}^n$, 
which essentially describes a manifold of horizon cross-section. 
We also define the covariant derivatives with respective to $g_{\mu \nu}, 
{}^{m}g_{ab}, \gamma_{ij}$, by $\nabla_\mu ,\: D_a, \: \hat{D}_i$, respectively. 
The non-vanishing components of the Christoffel 
symbol $\Gamma^\lambda{}_{\mu \nu}$ associated with $g_{\mu \nu}$ are given explicitly as  
\bena
 \Gamma^a{}_{bc} = \tilde \Gamma^a{}_{bc} \,, \quad 
  \Gamma^a{}_{ij} = -R(D^a R) \gamma_{ij} \,, \quad 
   \Gamma^i{}_{aj} = \frac{D_a R}{R}\delta^i{}_j \,, \quad 
    \Gamma^i{}_{jk} = \hat \Gamma^i{}_{jk} \,, \quad 
\label{def:Christoffel}
\eena
where $\tilde \Gamma^a{}_{bc}$ and $\hat \Gamma^i{}_{jk}$ are the components of the Christoffel symbols 
associated with the metrics ${}^{m}g_{ab}$ and $\gamma_{ij}$, respectively.

\subsection{Proca equations in the general warped product background}

Let us consider the massive vector field $A_\mu$ with mass-squared $\mu^2$ 
in $(M,g_{\mu \nu})$, which obeys the following Proca equation:  
\bena
 \nabla_\nu F^{\mu \nu} + \mu^2 A^\mu =0 \,, 
\label{eq:proca}
\eena
where the field strength $F$ is given as usual $F_{\mu \nu }:= \nabla_\mu A_\nu - \nabla_\nu A_\mu$. 
In addition, the following Lorenz condition needs to be satisfied;
\ben
\nabla_\mu A^\mu =0 \,.
\label{condi:lorenz}
\een
Since Proca equation, (\ref{eq:proca}), is not gauge-invariant due to the mass term, 
massive vector field in $D$-dimensions has $D-1$ physical degrees of freedom. 
Note that the Proca equation naturally arises, via Kaluza-Klein compactification,  
from linear gravitational perturbations in higher dimensional black holes, see, e.g. \cite{Ishibashi:Pani:Gualtieri:Cardoso:2015}.  

\medskip

By using the formulas (\ref{def:Christoffel}) above, we can express the projection of 
Proca equation (\ref{eq:proca}) on ${\cal N}^m$ and that of on ${\cal K}^n$, respectively, as
\bena 
&{}& D_bF^{ab} + n\frac{D_bR}{R} F^{ab} + {\hat D}_jF^{aj} + \mu^2 A^a =0   \,, 
\label{proca:comp:a}
\\ 
&{}& D_bF^{ib} + n\frac{D_bR}{R} F^{ib}  + {\hat D}_j F^{ij} + \mu^2 A^i = 0 \,,
\label{proca:comp:i} 
\eena
and the Lorenz condition (\ref{condi:lorenz}) as 
\ben
   D_aA^a + n\frac{D_aR}{R}A^a + {\hat D}_iA^i =0 \,.
\label{condi:lorenz:ai}
\een

\medskip
Now we discuss a decomposition of $A_\mu$. Note first that any dual vector field $v_i$ on ${\cal K}^n$ can be 
expressed as 
$$v_i= V_i + {\hat D}_i S \,,$$ 
where $\hat D^i V_i=0$, and $V_i$ and $S$ are called, respectively, the {\em vector-} and {\em scalar-}type 
components of $v_i$. 
Note that the vector-type is called sometime the {\em axial-} or {\em odd-}type components, and {scalar-}type is called {\em polar-} or {\em even-}type component. 
In the same manner, we can decompose any dual vector field $A_\mu$ in the background (\ref{def:background}) into the vector-type and scalar-type according to their tensorial behavior on ${\cal K}^n$. 
Namely, we can express $A_\mu$ as
\bena
A_\mu dx^\mu = A^S_a dy^a + {\hat D}_i A^S dz^i + A^V_i dz^i \,, \quad {\hat D}^iA^V_i =0 \,. 
\eena
We refer to $A^V_i$ as the vector-type and $A^S_a, \; A^S$ as the scalar-type components.  
  
\medskip

Next, let us introduce scalar harmonics ${\mS}_{{\bf k}_S}$ as
\ben
 ({\hat \triangle} + {k}_S^2 ) {\mS}_{{\bf k}_S  } = 0 \,, \quad \int_{\cal K}  d\sigma_n {\mS}_{{\bf k}'_S} {\mS}_{{\bf k}_S}
  = \delta_{ {{\bf k}'_S }  {{\bf k}_S } }  \,,  
\een
where ${\hat \triangle}:= \gamma^{ij}\hat D_i \hat D_j=\hat D^i \hat D_i$, and $d\sigma_n$ denotes the volume element on 
${\cal K}^n$. 
Note that when ${\cal K}^n$ is the unit $n$-sphere, the eigenvalue is given by ${k}_S^2= l(l+n-1) \,, \; l=0,1,2,\dots$. 
Similarly we introduce vector harmonics ${\mV}_{{\bf k}_V i}$ on ${\cal K}^n$ as 
\ben
 ({\hat \triangle} + {k}_V^2 ) {\mV}_{{\bf k}_V  i} = 0 \,, 
 \quad
        {\hat D}^i {\mV}_{{\bf k}_V i} = 0\,, 
 \quad 
 \int_{\cal K} d\sigma_n {\mV}_{{\bf k}'_V}^j {\mV}_{{\bf k}_V j}
  = \delta_{ {{\bf k}'_V }  {{\bf k}_V } }  \,, 
\een
where the eigenvalue is given, when ${\cal K}^n$ is the unit $n$-sphere, by 
${k}_V^2= l(l+n-1) - 1\,, \; l=1,2, \dots$. The number of independent components of 
$ {\mV}_{{\bf k}_V  i}$ is $n-1$ and only when $n \geqslant 2$, the odd-part is non-trivial. 

\medskip

We can expand the vector- and scalar-type components of $A_\mu$ in terms of the above scalar and 
vector harmonics: for vector-type, 
\ben
  A^V_i = \sum_{{\bf k} V} \phi_{{\bf k} V} {\mV}_{{\bf k}_V i}\,, 
\een
where $\phi_{{\bf k} V}(y)$ is a function on ${\cal N}^m$, 
and for scalar-type, 
\ben 
 A^S_a =  \sum_{{\bf k} S} A_{{\bf k} a} {\mS}_{{\bf k}_S} \,, 
\quad 
 A^S =  \sum_{{\bf k} S} A_{{\bf k} } {\mS}_{{\bf k}_S} \,,  
\een
where $A_{{\bf k} a}(y)$ and $A_{{\bf k} }(y) $ are, respectively, 
vector and scalar fields on ${\cal N}^m$. 
Hereafter, we omit the indices ${\bf k}_S$, ${\bf k}_V$ for brevity. 

\medskip

Now that we have separated the variables by introducing the scalar and vector harmonics, we can 
reduce the Proca equation, (\ref{eq:proca}), to a set of equations for $\phi^V$ and for $(A_a, A)$ 
on ${\cal N}^m$. 

\subsubsection{Vector-type component of the Proca equation}
The vector-type consists of a single scalar function $\phi^V$ on ${\cal N}^m$, and the field strength is written as   
\ben
 F^{ab}=0 \,, \quad F^{ai} = \frac{1}{R^2} (D^a \phi^V) {\mV}^i \,, \quad 
 F^{ij}= \frac{2}{R^4} \phi^V {\hat D}^{[i} {\mV}^{j]} \,. 
\een
It then immediately follows that the projection (\ref{proca:comp:a}) onto ${\cal N}^m$ and the Lorenz condition (\ref{condi:lorenz:ai}) trivially hold. The only non-trivial equation comes from (\ref{proca:comp:i}), which is written explicitly   
\ben
 {}^m \Box \phi^V + (n-2) \frac{D^aR}{R}D_a \phi^V - \left[ \frac{K(n-1) + k_V^2}{R^2}+\mu^2 \right]\phi^V
  = 0 \,, 
\label{eq:master:vect:m}
\een
where ${\hat D}_j {\hat D}^{i}{\mV}^{j}=K(n-1){\mV}^i$ has been used, and where here and hereafter 
${}^m \Box := D^aD_a$ is the d'Alembertian on the $m$-dimensional spacetime ${\cal N}^m$.  
This is the master equation for the vector-type component of the massive vector field $A_\mu$. 

\subsubsection{Scalar-type component of the Proca equation} 

For the scalar-type component, the field strength is given by 
\bena
  F^{ab} = 2D^{[a} B^{b]} {\mS} \,, \quad 
  F^{ai} = - \frac{1}{R^2}B^a {\hat D}^i{\mS} \,, \quad 
  F^{ij} = 0\,,  
\eena
where we have introduced 
\ben
 B^a := A^a -D^aA \,. 
\een
Then, the projections (\ref{proca:comp:a}) on ${\cal N}^m$,  (\ref{proca:comp:i}) on ${\cal K}^n$, 
and the Lorenz condition (\ref{condi:lorenz}) reduce, respectively, to
\bena
&{}& 2D_bD^{[a}B^{b]} + 2n\frac{D_bR}{R}D^{[a}B^{b]}+ \left(\frac{k_S^2}{R^2}+\mu^2 \right)B^a
      + \mu^2 D^a A =0 \,, 
\label{eq:scal:a}
\\
&{}& D_bB^b + (n-2)\frac{D_bR}{R}B^b + \mu^2 A = 0 \,, 
\label{eq:scal:i}
\\
&{}& D_bB^b + n\frac{D_bR}{R}B^b + {}^m \Box A + n\frac{D_bR}{R} D^bA - \frac{k_S^2}{R^2}A = 0 \,.
\label{eq:scal:lorenz}
\eena
Acting $D^a$ on (\ref{eq:scal:i}) and then combining with (\ref{eq:scal:a}), 
we can obtain the equation only for $B^a$ as  
\ben
 {}^m \Box B^a - {}^m {\cal R}^a{}_b B^b + \frac{D_bR}{R}\left(nD^bB^a -2D^aB^b \right)
         + (n-2)D^a\left(\frac{D_bR}{R}\right) B^b 
         - \left(\frac{k_S^2}{R^2}+\mu^2 \right) B^a = 0 \,,  
\label{eq:Ba}
\een
where ${}^m {\cal R}_{ab}$ is the Ricci tensor on ${\cal N}^m$, while 
combining (\ref{eq:scal:i}) and (\ref{eq:scal:lorenz}) we have 
\ben
 {}^m \Box A + n\frac{D_cR}{R}D^cA - \left(\frac{k_S^2}{R^2}+\mu^2 \right) A 
  + 2\frac{D_bR}{R}B^b =0 \,.
\label{eq:A} 
\een
Due to the last term in (\ref{eq:A}), the scalar variable $A$ is coupled with $B_a$. 
Inspecting eqs.~(\ref{eq:Ba}) and (\ref{eq:A}), we can expect to be able 
to obtain a set of decoupled equations when $R=const. $, as clearly the case for $A$ in (\ref{eq:A}). 
In the next section, we shall show that this is also the case for $B_a$; One can in fact derive from 
(\ref{eq:Ba}) a single master equation for a single component of $B_a$ when considering, as 
our background geometry, the near-horizon geometry of (near-)extremal black holes. 

\medskip 
Note that the scalar-type components $(B_a, A)$ together with the Lorenz condition 
describe $m$ dynamical degrees of freedom, 
while the vector-type components (though including only a single scalar field $\phi^V$) 
describe $n-1$ dynamical degrees of freedom as the vector harmonics ${\mV}_i$ itself has $n-1$ independent components. Thus, in total $m+n-1=D-1$ degrees of freedom for the massive vector 
field can be expressed by the above variables, as should be so. 

\subsection{The massless vector (Maxwell) field in the warped product background}
Before going further, we show that for the massless vector field case, 
one can in fact obtain a single master equation also for the scalar-type component. 
Let us consider the case $m=2$. When the mass vanishes $\mu^2=0$, eq.~(\ref{eq:scal:i}) reduces to 
\ben
  D_b \left( R^{n-2} B^b \right) =0 \,.
\een
This implies that there exists a scalar field $\phi^S$ on ${\cal N}^2$ such that
\ben
 D_a \phi^S = \epsilon_{ab} R^{n-2} B^b \,, 
\een
where $\epsilon_{ab}$ denotes the natural volume element on $({\cal N}^2, {}^2g_{ab})$. 
Since $\mu^2=0$, the gauge-invariance is recovered and $\phi^S$ admits a gauge-freedom. 
Now, as $\epsilon^{ca}\epsilon_{ab}= \delta^c{}_b$, it follows 
\ben
 F^{ai} = - \frac{1}{R^n} \epsilon^{ac}(D_{c} \phi^S) {\hat D}^i {\mS} \,.
\een
That $F^{ai}$ itself is gauge-invariant implies that the gauge freedom of $\phi^S$ is restricted to 
the replacement;  
\ben
 \phi^S \rightarrow \phi^S + const. \,. 
\label{gauge:res}
\een
In terms of $\phi^S$, eq.~(\ref{proca:comp:a}) is expressed as 
\ben
 \epsilon^{ab}D_b \left[R^n D^c\left(\frac{D_c \phi^S}{R^{n-2}}\right)-k_S^2 \phi^S \right]=0 \,.
\een
Therefore we have
\ben
 R^n D^c\left(\frac{D_c \phi^S}{R^{n-2}}\right)-k_S^2 \phi^S = c \,, 
\een
with $c$ being an arbitrary constant. We can always absorb this 
integration constant $c$ in $\phi^S$ by using the remaining gauge freedom (\ref{gauge:res}) and thus obtain 
\ben
  R^{n-2} D^c\left(\frac{D_c \phi^S}{R^{n-2}}\right) - \frac{k_S^2}{R^2} \phi^S = 0 \,.
\label{eq:master:maxwell}
\een
This is the master equation for the scalar-type component of Maxwell field, which is responsible for 
only a single polarization degree of freedom. Note that the master equation for the vector-type component of Maxwell field, 
given by eq.~(\ref{eq:master:vect:m}) with $m=2$ and the vanishing mass $\mu=0$ describes 
$n-1$ degrees of freedom as vector harmonics ${\mV}_i$ has $n-1$ independent components. Thus, in total, all $n=D-2$ independent degrees of freedom 
for the Maxwell field in $D=2+n$ dimensions can be expressed by the two master variables 
$\phi^V$ and $\phi^S$ together with vector and scalar harmonics $\mV_i$ and $\mS$. 

\medskip 

Note also that all the results obtained above hold in a fairly generic class of 
background spacetimes as far as they have the warped product structure given by eq.~(\ref{def:background}). In particular, the background geometry used so far is neither required to be a solution to the Einstein equations with any specific energy-momentum tensor, 
nor to possess any symmetry. 

\section{Extremal and near-extremal black holes and their near-horizon geometries}
\label{sect:Extremal}
In this section, we discuss the Proca equation when our background spacetime describes extremal or near-extremal black holes.  

\medskip 

From now on we assume $m=2$. Then the metric (\ref{def:background}) includes 
the standard solutions to the Einstein-Maxwell-$\Lambda$ system with 
$\Lambda$ being a cosmological constant when $m=2$, $y^a=(t,r)$, $R(y)=r$ and 
\bena
 {}^{2}g_{ab}dy^a dy^b = -F(r)dt^2 + \frac{ dr^2}{F(r)} \,, 
  \quad 
  F(r):= K - \frac{2M}{r^{n-1}} + \frac{Q^2}{r^{2(n-1)}}- \frac{2\Lambda}{n(n+1)} r^2 \,,
\label{def:metric:2}
\eena
where $M$ and $Q$ are, respectively, the mass and charge parameters. 
The black hole horizon is located at $r=r_+$ for which $F(r_+)=0$. In particular, 
the above metric allows for two horizons (or more) and then possesses a limit wherein 
the horizon becomes degenerate. Such a black hole is called extremal. The most well-known 
case is the Reissner-Nordstrom metric in four-dimensions, for which $n=2$, $K=1$, $\Lambda =0$, so that 
\ben
 F(r) = \frac{(r-r_+)(r-r_-)}{r^2} \,, \quad r_\pm := M \pm \sqrt{M^2-Q^2} \,. 
\een 
The extremal limit is the case $r_+=r_-=M=|Q|$. 
Even for the neutral (no electric charge $Q=0$) case, (\ref{def:metric:2}) admits 
an extremal black hole when $K=-1$, $\Lambda<0$, and $M<0$.   

\medskip 

We consider the case in which there are two horizons at $r_+$ and $r_-$. 
Since we are concerned with the neighborhood of the black hole (outer) horizon, instead of 
the Schwarzschild type coordinates used in (\ref{def:metric:2}), let us take the ingoing
Eddington-Finkelstein type coordinates, which cover the black hole horizon and 
in which our background metric (\ref{def:background}) takes the form:
\bena
 ds^2 &=& - F(r) dv^2 + 2dv dr + R(r)^2 \gamma_{ij}dz^idz^j \,, 
\non 
\\
 &{}& 
\non 
\\
 &{}& F(r) = (r-r_+)(r-r_-)g(r) \,, 
\label{def:metric:ef}
\eena  
where $g(r)>0$ is an everywhere regular function, except at true curvature 
singularity, and $v$ is the advanced time-coordinate. 
We introduce the extremality parameter as
\ben
\sigma := \frac{r_+ - r_-}{r_+} \,.
\label{def:param:extremal}
\een
When $\sigma \ll 1$, we refer to the metric (\ref{def:metric:ef}) as 
the {\em near-extremal} and when $\sigma=0$, {\em extremal}. 
For convenience, we also introduce the new radial coordinate 
\ben
 x:= \frac{r-r_+}{r_+} \,, 
\een
so that the black hole event horizon is located at $x=0$. 

\medskip 

It is known that any extremal black hole admits a near-horizon limit. 
Let us take the following scaling transformation:  
\ben
 x \rightarrow \lambda x \,, \quad 
 v \rightarrow \frac{r_+}{\lambda} v \,, \quad
 R \rightarrow r_+ R \,, \quad 
 \sigma \rightarrow \lambda \sigma \,,
\label{def:scaling}    
\een
with the scaling parameter $\lambda > 0$.  
Then the metric (\ref{def:metric:ef}) is written as,   
\bena
 \frac{ds^2}{r_+^2}&=&
             - F(\lambda x)dv^2 + 2dvdx +  R(\lambda x)^2 \gamma_{ij}dz^idz^j \,,
\non 
\\
    &{}& \qquad F(\lambda x) = x(x+ \sigma) g(\lambda x) \,.
\label{def:metric:lambda}
\eena 

The limiting $\lambda \rightarrow 0$ metric is the {\em near-horizon geometry},  
in which in particular $R$ becomes a constant, and the isometry is in general enhanced to be $O(2,1)$ as shown 
in \cite{KLR07}. 
Note that at this point, the scaling transformation (\ref{def:scaling}) is simply 
a change of the coordinates together with the parameter change, and the above 
metric (\ref{def:metric:lambda}) satisfies the same Einstein equations 
which the original metric satisfies. This is, however, not necessarily the case 
when we take a power expansion of the above metric by $\lambda$ 
and truncate at some order.    

\section{Expanding massive vector field and (near-) extremal black hole geometry} 
\label{sect:Expanding}

Now we shall develop our perturbation method. 
We view the scaling parameter $\lambda$ as the small perturbation parameter
 and consider a one-parameter family of the massive vector field 
$A_\mu(\lambda)$. We expand it in a power series in $\lambda$ about $\lambda=0$, as
$$
A_\mu(\lambda)=A^{(0)}_\mu + \lambda A^{(1)}_\mu + \lambda^2 A^{(2)}_\mu + \cdots \,.
$$ 
We also consider our background (near-)extremal black hole metric 
as a one parameter family of metrics, expanded as 
$$
g_{\mu \nu}(\lambda)=g^{(0)}_{\mu \nu} 
            + \lambda g^{(1)}_{\mu \nu} + \lambda^2g^{(2)}_{\mu \nu}+\cdots \,,
$$
with the leading metric $g_{\mu \nu}^{(0)}=g_{\mu \nu}|_{\lambda =0}$ being the 
corresponding near-horizon geometry. 
Note however that the background geometry is fixed from the 
beginning, and in particular not required to solve the Einstein equations at 
each order. Then, in doubly expanding the field variables and the background 
metric with respect to $\lambda$ and exploiting the enhanced symmetry of the 
leading-order near-horizon geometry, we will examine the Proca equation 
at each order of $\lambda$. 
We will perform this analysis in the vector- and scalar-type components, separately. 
  
\subsection{Proca equations in (near-)extremal Reissner-Nordstrom black hole}

For concreteness, we first consider the standard four-dimensional 
Reissner-Nordstrom black hole background case. Generalisation of our 
formulas to more generic case is given afterward. The metric of the Reissner-Nordstrom black hole is given by $m=n=2$, $K=1$, 
(with the scaling (\ref{def:scaling}), as 
\ben
\frac{ds^{2}}{r_+^2}= -F(x)dv^{2}+2dvdx+ {R^{2}} d\Omega^{2} \,,
\label{eq:(vx)metric}
\een
where 
\ben
F(x) = \frac{x(x+\sigma)}{(1+\lambda x)^2} \,, \quad  R= (1+\lambda x) \,.
\label{eq:RN_F_R}
\een
%

\subsubsection{Vector-type component of the Proca equation}
We begin with the vector-type component. In the Reissner-Nordstrom background 
(\ref{eq:(vx)metric}), the Proca equation for the vector-type, (\ref{eq:master:vect:m}), reduces to 
%
\ben
\left[F\partial_{x}^{2}+ (\partial_x F) \partial_{x}+2\partial_{v}\partial_{x}-\left\{\frac{k^{2}_{V}+1}{(1+\lambda x)^{2}}+\mu^{2} r_{+}^{2}\right\}\right]\phi^V=0 \,.
%
\label{eq:Proca_vector2}
\een

\medskip 

Let us expand the master scalar variable $\phi^V$ in a power series of $\lambda$ as 
\ben
 \phi^V =\sum^{\infty}_{n=0}\lambda^{n}\Phi_V^{(n)} \,.
\label{eq:phi_expand}
\een
Pluging this into (\ref{eq:Proca_vector2}), as well as expanding $F$ and the other 
$\lambda$-dependent coefficients in (\ref{eq:Proca_vector2}), we obtain  
\ben
\sum^{\infty}_{n=0}\lambda^{n}\left[\sum^{n}_{m=0}L^{(m)}_{V} \Phi_V^{(n-m)}\right]=0 \,,
\label{eq:Proca_vec_expand}
\een
where we have introduced the following series of differential operators on 
${\cal N}^2$, 
\bena
L^{(n)}_{V} &:=& (-1)^{n} \Big[
                                       (n+1) x^{n+1}(x+\sigma)\partial_{x}^{2}
                                     +(n+1)x^{n} [(n+2)x+(n+1)\sigma]  \partial_{x} 
\non \\
              &{}&   \qquad \qquad  
                                    +2\delta_{n0}\partial_{v}\partial_{x}
                                     -(n+1)(k^{2}_{V}+1)x^{n}-\delta_{n0}\mu^{2}r_{+}^{2}
                            \Big] \,, 
%
\label{eq:phi_op}
\eena 
where here and hereafter $\delta_{ij}$ denotes the kronecker's delta. 
Therefore, at each order 
\ben
\sum^{n}_{m=0}L^{(m)}_{V}\Phi_V^{(n-m)}=0 \,.
\label{eq:Proca_vec_expand2}
\een
Namely, we have: 
\bena
L^{(0)}_{V}\Phi_V^{(0)}&=&0 \,, 
\label{eq:v:0}
\\
L^{(0)}_{V}\Phi_V^{(1)}&=&-L^{(1)}_{V}\Phi^{(0)} \,, 
\label{eq:v:1}
\\
L^{(0)}_{V}\Phi_V^{(2)}&=&-L^{(1)}_{V}\Phi^{(1)}-L^{(2)}_{V}\Phi_V^{(0)} \,, 
\label{eq:v:2}
\\
&\vdots&\non
\\
L^{(0)}_{V}\Phi_V^{(n)}&=&-\sum^{n}_{m=1}L^{(m)}_{V} \Phi_V^{(n-m)} \,. 
\label{eq:v:n}
\eena
The leading order master equation~(\ref{eq:v:0}) is homogeneous, and at each sub-leading 
inhomogeneous equation has a source term that consists only of the lower-order 
variables. Thus, once having obtained the leading solution $\Phi_V^{(0)}$, one can 
successively obtain all order solutions $\Phi_V^{(n)}$. 

\medskip 

Here we give the general solution to the leading-order master equation~(\ref{eq:v:0}): $L_V^{(0)} \Phi_V^{(0)}=0$. 
With the ansatz of time-dependency $\Phi \propto e^{-i\omega v}$, the time-derivative $\partial_v$ is replaced with 
$-i\omega$ and the leading-order operator is rewritten as the second-order ordinary differential operator as 
\ben
 L_V^{(0)} = x(x+\sigma) \frac{d^2}{dx^2} + (2x+ \sigma -2 i \omega) \frac{d}{dx} - (k_V^2 + 1 + \mu^2r_+^2) \,.
\een
Note that the time-coordinate $v$ is scale-transformed as $v \rightarrow (r_+ /\lambda) v $ in eq.~(\ref{def:scaling}), and accordingly the frequency $\omega$ is also scale-transformed as $\omega \rightarrow (\lambda/r_+) \omega$. 
The general solution is then, for the near-extremal $\sigma \neq 0$ case:
\bena
\Phi_V^{(0)}
 &=& C_1\cdot {}_2F_1\left( - \nu + \frac{1}{2}, \nu+\frac{1}{2}, 1+ 2i \frac{\omega}{\sigma}; 1 + \frac{x}{\sigma} \right) 
 \non \\
  &{}&  
     + C_2 \cdot (x+ \sigma)^{-2i \omega/\sigma} 
             {}_2F_1\left( - \nu + \frac{1}{2}- 2i \frac{\omega}{\sigma}, \nu+\frac{1}{2}- 2i \frac{\omega}{\sigma}, 1- 2i \frac{\omega}{\sigma}; 1 + \frac{x}{\sigma}  \right) \,, 
\eena
where ${}_2F_1$ denotes the hypergeometric function and 
\bena
  \nu := \sqrt{k_V^2+1 +\mu^2r_+^2 + \frac{1}{4}} \,,  
\eena
and where $C_1$ and $C_2$ are arbitrary constants. 
As for the extremal $\sigma=0$ case:
\bena
 \Phi_V^{(0)}
 &=&  \frac{1}{\sqrt{x}}e^{-i\omega/x}
   \left[
           C_1\cdot I_\nu \left( i \omega/x \right) +   C_2\cdot K_\nu \left( i \omega/x \right) 
   \right] \,,
\eena
where $I_\nu, \:K_\nu$ denote the modified Bessel functions. 
Then, once boundary conditions of interest are determined, one can 
construct the Green's function $G_V^{(0)} = L_V^{(0)}{}^{-1}$ by standard argument, and obtain higher order solutions, which are formally expressed as
\ben
\Phi_V^{(n)} = -G_V^{(0)}  \sum^{n}_{m=1}L^{(m)}_{V} \Phi_V^{(n-m)} \,.  
\een

\subsubsection{Scalar type component of the Proca equation}
Let us turn to the scalar-type components of Proca equation, (\ref{eq:Ba}) and (\ref{eq:A}), 
in the four-dimensional Reissner-Nordstrom background case. 
Setting $m=n=2, K=1, \Lambda=0$, we have %
\ben
D_{c}D^{c}A+2\left(\frac{D_{a}R}{R}\right)(D^{a}A)-\left(\frac{k^{2}_{S}}{R^{2}}+\mu^{2}\right)A+2\left(\frac{D_{a}R}{R}\right)B^{a}=0 \,, 
\label{eq:Proca_scalar1}
\een
\ben
-D_{c}D^{c}B_{a}+{}^{2}{\cal R}_{a}{}^{c}B_{c}+4\left(\frac{D^{b}R}{R}\right)D_{[a}B_{b]}+\left(\frac{k^{2}_{S}}{R^{2}}+\mu^{2}\right)B_{a}=0 \,, 
\label{eq:Proca_scalar2}
\een
where ${}^2{\cal R}^a{}_b$ is the Ricci tensor on ${\cal N}^2$, given in terms of the present coordinates $y^a=(v,r)$ by 
\ben
{}^{2}{\cal R}_{a}{}^{b}=-\frac{(\partial_{x}^{2}F)}{2r_{+}^{2}}\delta_{a}{}^{b} \,.
\een
In the coordinate system of (\ref{eq:(vx)metric}), the above equations, (\ref{eq:Proca_scalar1}) 
and  (\ref{eq:Proca_scalar2}), are explicitly written as the coupled equations for three-components, $(A,B_x,B_v)$, 
\bena
&& \left[
        F\partial_{x}^{2}+(\partial_{x}F)\partial_{x}
       +2\partial_{v}\partial_{x} + \frac{2\lambda}{1+\lambda x}(F\partial_{x}+\partial_{v}) 
       -\left\{\frac{k^{2}_{S}}{(1+\lambda x)^{2}}+\mu^{2} r_{+}^{2}\right\} 
       \right]A
\non        
\\
&& \hspace{8cm}
+\frac{2\lambda}{1+\lambda x} \left( FB_{x}+B_{v} \right)=0 \,,
%
\label{eq:Proca_scalar1_ver2}
\eena
\bena
&& \left[ F\partial_{x}^{2}+2(\partial_{x}F)\partial_{x}
          +(\partial_{x}^{2}F)\partial_{x}
          +2\partial_{v}\partial_{x}+\frac{2\lambda}{1+\lambda x}\partial_{v}
          -\left\{
                     \frac{k^{2}_{S}}{(1+\lambda x)^{2}}+\mu^{2} r_{+}^{2} 
           \right\}
   \right]B_{x}
\non 
\\
&& \hspace{10cm} 
 -\frac{2\lambda}{1+\lambda x}\partial_{x}B_{v}=0 \,,
%
\label{eq:Proca_scalar2_ver2-1}
\eena
\bena
&& \left[ 
            F\partial_{x}^{2}+2\partial_{v}\partial_{x}
           + \frac{2\lambda}{1+\lambda x}F\partial_{x}
           - \left\{
                       \frac{k^{2}_{S}}{(1+\lambda x)^{2}} + \mu^{2} r_{+}^{2} 
              \right\} 
     \right]B_{v} 
\non 
\\
&{}& \hspace{7cm} 
     +\left[
              (\partial_{x}F)\partial_{v} -\frac{2\lambda}{1+\lambda x}F\partial_{v}
      \right]B_{x}=0 \,.
%
\label{eq:Proca_scalar2_ver2-2}
\eena

Now we expand the variables $(A,\: B_x, \:B_v)$ about $\lambda$ as 
\ben
A=\sum_{n=0}^{\infty} \lambda^{n}\Phi_{S1}^{(n)} \,, 
\label{eq:A_expand}
\een
\ben
B_{x}=\sum_{n=0}^{\infty} \lambda^{n}\Phi_{S2}^{(n)} \,, 
\label{eq:Bx_expand}
\een
\ben
B_{v}=\sum_{n=0}^{\infty} \lambda^{n}\Phi_{S3}^{(n)} \,.
\label{eq:Bv_expand}
\een
Also we expand each term appearing in eqs.~(\ref{eq:Proca_scalar1_ver2}), 
(\ref{eq:Proca_scalar2_ver2-1}), and (\ref{eq:Proca_scalar2_ver2-2}) about $\lambda$. 
%
In order to express the equations (\ref{eq:Proca_scalar1_ver2}), (\ref{eq:Proca_scalar2_ver2-1}), and (\ref{eq:Proca_scalar2_ver2-2}) at each order, it is convenient to introduce the following set of differential operators: 
\bena
L^{(n)}_{\alpha 1} &:=&  (-1)^{n}[(n+1) x^{n+1}(x+\sigma)\partial_{x}^{2}+\{(n+1)x^{n}(2x+\sigma) +2\delta_{n0}\partial_{v}\}\partial_{x}
\non \\
&{}& \qquad  \quad 
 +2(\delta_{n0}-1)x^{n-1}\partial_{v}-(n+1)k^{2}_{S}x^{n} -\delta_{n0}\mu^{2}r_{+}^{2}] \,, 
\label{eq:aA_op}
\\
&{}& \non \\
L^{(n)}_{\alpha 2} &:=&  (-1)^{n+1}n(n+1) x^{n}(x+\sigma) \,, 
\label{eq:ax_op}
\\
&{}& \non \\
L^{(n)}_{\alpha 3} &:=& 2(-1)^{n}(\delta_{n0}-1)x^{n-1} \,, 
\label{eq:av_op}
\\ 
&{}& \non \\
L^{(n)}_{\beta 2} &:=&  (-1)^{n}[(n+1) x^{n+1}(x+\sigma)\partial_{x}^{2}
\non \\
&{}& \qquad \quad
  +2(n+1)x^{n}\{(n+2)x+(n+1)\sigma\}\partial_{x}
\non \\
&{}& \qquad \quad
     +2\delta_{n0}\partial_{v}\partial_{x}+(n+1)^{2}x^{n-1}\{(n+2)x+n\sigma\}
\non \\
&{}& \qquad \quad
 +2(\delta_{n0}-1)x^{n-1}\partial_{v}-(n+1)k^{2}_{S}x^{n}
-\delta_{n0}\mu^{2}r_{+}^{2}] \,,
\label{eq:bx_op}
\\
&{}& \non \\
L^{(n)}_{\beta 3} &:=& 2(-1)^{n}(1-\delta_{n0})x^{n-1}\partial_{x} \,,
\label{eq:bv_op}
\\
&{}& \non \\
L^{(n)}_{\gamma 2} &:=&  (-1)^{n}(n+1)x^{n}\{2(n+1)x+(2n+1)\sigma\}\partial_{x} \,, 
\label{eq:cx_op}
\\
&{}& \non \\ 
 L^{(n)}_{\gamma 3} &:=& (-1)^{n}[(n+1) x^{n+1}(x+\sigma)\partial_{x}^{2}-n(n+1)x^{n}(x+\sigma)\partial_{x}
 \non \\
 &{}& \qquad \quad
 +2\delta_{n0}\partial_{v}\partial_{x}-(n+1)k^{2}_{S}x^{n}
-\delta_{n0}\mu^{2}r_{+}^{2}] \,. 
\label{eq:cv_op}
\eena
In terms of these operators, 
eqs.~(\ref{eq:Proca_scalar1_ver2}), (\ref{eq:Proca_scalar2_ver2-1}), 
and (\ref{eq:Proca_scalar2_ver2-2}), are expressed as 
\ben
\sum^{\infty}_{n=0}\lambda^{n}\left[\sum^{n}_{m=0}\left\{L^{(m)}_{\alpha 1}\Phi_{S1}^{(n-m)}+L^{(m)}_{\alpha 2}\Phi_{S2}^{(n-m)}+L^{(m)}_{\alpha 3}\Phi_{S3}^{(n-m)}\right\}\right]=0 \,,
\label{eq:Proca_scalar1_ver3}
\een
\ben
\sum^{\infty}_{n=0}\lambda^{n}\left[\sum^{n}_{m=0}\left\{L^{(m)}_{\beta 2}\Phi_{S2}^{(n-m)}+L^{(m)}_{\beta 3}\Phi_{S3}^{(n-m)}\right\}\right]=0
\,,
~~~~~~~~~~~~~~~~~~~
\label{eq:Proca_scalar2_ver3-1}
\een
\ben
\sum^{\infty}_{n=0}\lambda^{n}\left[\sum^{n}_{m=0}\left\{L^{(m)}_{\gamma 2}\Phi_{S2}^{(n-m)}+L^{(m)}_{\gamma 3}\Phi_{S3}^{(n-m)}\right\}\right]=0 \,. 
~~~~~~~~~~~~~~~~~~~
\label{eq:Proca_scalar2_ver3-2} 
\een
Therefore we have, at each order, the following equations: 
\ben
\sum^{n}_{m=0}\left\{L^{(m)}_{\alpha 1}\Phi_{S1}^{(n-m)}+L^{(m)}_{\alpha 2}\Phi_{S2}^{(n-m)}+L^{(m)}_{\alpha 3}\Phi_{S3}^{(n-m)}\right\}=0 \,,
\label{eq:Proca_scalar1_ver4}
\een
\ben
\sum^{n}_{m=0}\left\{L^{(m)}_{\beta 2}\Phi_{S2}^{(n-m)}+L^{(m)}_{\beta 3}\Phi_{S3}^{(n-m)}\right\}=0 \,,
~~~~~~~~~~~~~~~~~~~
\label{eq:Proca_scalar2_ver4-1}
\een
\ben
\sum^{n}_{m=0}\left\{L^{(m)}_{\gamma 2}\Phi_{S2}^{(n-m)}+L^{(m)}_{\gamma 3}\Phi_{S3}^{(n-m)}\right\}=0 \,.
~~~~~~~~~~~~~~~~~~~
\label{eq:Proca_scalar2_ver4-2}
\een
These equations can be rewritten in the following manner:
\begin{itemize}
\item[(i)] At the leading order $\lambda =0$, the geometry is the near-horizon 
geometry, and we find  
%
\ben
L^{(0)}_{\alpha 2}=L^{(0)}_{\alpha 3}=L^{(0)}_{\beta 3}=0 \,. 
\een
Therefore we have 
\ben
\left(
\begin{tabular}{ccc}
$L^{(0)}_{\alpha 1}$&$0$&$0$\\
$0$&$L^{(0)}_{\beta 2}$&$0$\\
$0$&$L^{(0)}_{\gamma 2}$&$L^{(0)}_{\gamma 3}$\\
\end{tabular}
\right)
\left(
\begin{tabular}{c}
$\Phi_{S1}^{(0)}$\\
$\Phi_{S2}^{(0)}$\\
$\Phi_{S3}^{(0)}$\\
\end{tabular}
\right)
=
\left(
\begin{tabular}{c}
$0$\\
$0$\\
$0$\\
\end{tabular}
\right) \,.
\label{eq:Proca_scalar_0}
\een
This shows that the first two equations are mutually decoupled, homogeneous master equations for the two master variables, $\Phi_{S1}^{(0)}, \: \Phi_{S2}^{(0)}$. 
These two variables describe two dynamical degrees of freedom, which 
the scalar-type components should be responsible for describing. 
(Recall that in four-dimensions, the massive vector field has in total 
three dynamical degrees of freedom, one of which is expressed by the vector-type 
component.)  
By using the third equation, the remaining variable $\Phi_{S3}^{(0)}$ can 
be determined in terms of $\Phi_{S2}^{(0)}$.  

\item[(ii)]
Next, at the first order of $\lambda$, we have 
\ben
\left(
\begin{tabular}{ccc}
$L^{(0)}_{\alpha 1}$&$0$&$0$\\
$0$&$L^{(0)}_{\beta 2}$&$0$\\
$0$&$L^{(0)}_{\gamma 2}$&$L^{(0)}_{\gamma 3}$\\
\end{tabular}
\right)
\left(
\begin{tabular}{c}
$\Phi_{S1}^{(1)}$\\
$\Phi_{S2}^{(1)}$\\
$\Phi_{S3}^{(1)}$\\
\end{tabular}
\right)
=
-
\left(
\begin{tabular}{ccc}
$L^{(1)}_{\alpha 1}$&$L^{(1)}_{\alpha 2}$&$L^{(1)}_{\alpha 3}$\\
$0$&$L^{(1)}_{\beta 2}$&$L^{(1)}_{\beta 3}$\\
$0$&$L^{(1)}_{\gamma 2}$&$L^{(1)}_{\gamma 3}$\\
\end{tabular}
\right)
\left(
\begin{tabular}{c}
$\Phi_{S1}^{(0)}$\\
$\Phi_{S2}^{(0)}$\\
$\Phi_{S3}^{(0)}$\\
\end{tabular}
\right) \,.
\label{eq:Proca_scalar_1}
\een 
The first two equations are mutually decoupled, 
inhomogeneous wave equations for the two scalar variables, $\Phi_{S1}^{(1)}, \: \Phi_{S2}^{(1)}$, 
and via the third equation, the remaining variable $\Phi_{S3}^{(1)}$ can 
be determined. The source terms of the inhomogeneous wave equations consist 
of the leading order solutions $\Phi_{S1}^{(0)}, \: \Phi_{S2}^{(0)}, \: \Phi_{S3}^{(0)}$.    

\item[(iii)] At the second-order, we find 
\bena
\left(
\begin{tabular}{ccc}
$L^{(0)}_{\alpha 1}$&$0$&$0$\\
$0$&$L^{(0)}_{\beta 2}$&$0$\\
$0$&$L^{(0)}_{\gamma 2}$&$L^{(0)}_{\gamma 3}$\\
\end{tabular}
\right)
\left(
\begin{tabular}{c}
$\Phi_{S1}^{(2)}$\\
$\Phi_{S2}^{(2)}$\\
$\Phi_{S3}^{(2)}$\\
\end{tabular}
\right)
&=&-
\left(
\begin{tabular}{ccc}
$L^{(1)}_{\alpha 1}$&$L^{(1)}_{\alpha 2}$&$L^{(1)}_{\alpha 3}$\\
$0$&$L^{(1)}_{\beta 2}$&$L^{(1)}_{\beta 3}$\\
$0$&$L^{(1)}_{\gamma 2}$&$L^{(1)}_{\gamma 3}$\\
\end{tabular}
\right)
\left(
\begin{tabular}{c}
$\Phi_{S1}^{(1)}$\\
$\Phi_{S2}^{(1)}$\\
$\Phi_{S3}^{(1)}$\\
\end{tabular}
\right)
\non
\\
&{}&
-
\left(
\begin{tabular}{ccc}
$L^{(2)}_{\alpha 1}$&$L^{(2)}_{\alpha 2}$&$L^{(2)}_{\alpha 3}$\\
$0$&$L^{(2)}_{\beta 2}$&$L^{(2)}_{\beta 3}$\\
$0$&$L^{(2)}_{\gamma 2}$&$L^{(2)}_{\gamma 3}$\\
\end{tabular}
\right)
\left(
\begin{tabular}{c}
$\Phi_{S1}^{(0)}$\\
$\Phi_{S2}^{(0)}$\\
$\Phi_{S3}^{(0)}$\\
\end{tabular}
\right) \,.
\label{eq:Proca_scalar_2}
\eena
The structure of these second-order equations are the same as the first-order equations: The first two equations are mutually decoupled inhomogeneous wave equations for the two scalar variables $\Phi_{S1}^{(2)}, \: \Phi_{S2}^{(2)}$, and the third-equation is used to determine the remaining third variable $\Phi_{S3}^{(2)}$. The source terms for the inhomogeneous equations are given only in terms of the lower-order, i.e., the first- or the leading-order variables.

\item[(iv)] In general, at the $n$-th order, we have:  
\ben
\left(
\begin{tabular}{ccc}
$L^{(0)}_{\alpha 1}$&$0$&$0$\\
$0$&$L^{(0)}_{\beta 2}$&$0$\\
$0$&$L^{(0)}_{\gamma 2}$&$L^{(0)}_{\gamma 3}$\\
\end{tabular}
\right)
\left(
\begin{tabular}{c}
$\Phi_{S1}^{(n)}$\\
$\Phi_{S2}^{(n)}$\\
$\Phi_{S3}^{(n)}$\\
\end{tabular}
\right)
=
-
\sum_{m=1}^{n}
\left[
\left(
\begin{tabular}{ccc}
$L^{(m)}_{\alpha 1}$&$L^{(m)}_{\alpha 2}$&$L^{(m)}_{\alpha 3}$\\
$0$&$L^{(m)}_{\beta 2}$&$L^{(m)}_{\beta 3}$\\
$0$&$L^{(m)}_{\gamma 2}$&$L^{(m)}_{\gamma 3}$\\
\end{tabular}
\right)
\left(
\begin{tabular}{c}
$\Phi_{S1}^{(n-m)}$\\
$\Phi_{S2}^{(n-m)}$\\
$\Phi_{S3}^{(n-m)}$\\
\end{tabular}
\right)\right] \,. 
\label{eq:Proca_scalar_n}
\een
\end{itemize}
Thus, at any order, we find the same structure; 
we obtain two mutually decoupled master equations for two 
master variables, $\Phi_{S1}^{(n)}, \: \Phi_{S2}^{(n)}$, given respectively by the operators 
$L^{(0)}_{\alpha 1}, \; L^{(0)}_{\beta 2}$, and the third-equation is used to 
determine the remaining third variable $\Phi_{S3}^{(n)}$. 
The source terms for the inhomogeneous master equations are given 
by the lower-order variables. This is the set of master equations for 
the scalar-type components. Once having obtained the leading solutions 
$\Phi_{S1}^{(0)},\: \Phi_{S2}^{(0)}$, one can solve successively any order of  
the scalar-type components of the Proca equation.     

\medskip 

As in the vector-type case, one can immediately find the general solutions to the leading-order master equations, $L_{\alpha 1}^{(0)} \Phi_{S1}^{(0)}=0$ and 
$L_{\beta 2}^{(0)} \Phi_{S2}^{(0)}=0$. 
With the ansatz of time-dependency $\Phi \propto e^{-i\omega v}$, the time-derivative $\partial_v$ is replaced with 
$-i\omega$ and the two leading operators are, respectively, expressed as  
\bena
 L_{\alpha 1}^{(0)} &=& x(x+\sigma) \frac{d^2}{dx^2} + \left( 2x+ \sigma -2 i \omega \right) \frac{d}{dx} - (k_S^2 + \mu^2r_+^2) \,, 
 \\
  L_{\beta 2}^{(0)} &=& x(x+\sigma) \frac{d^2}{dx^2} +2 (2x+ \sigma - i \omega) \frac{d}{dx} - (k_S^2 + \mu^2r_+^2 -2 ) \,.
\eena 
Again note that according to the time-coordinate scaling $v \rightarrow (r_+ /\lambda) v $, 
the frequency is also scale-transformed: $\omega \rightarrow (\lambda/r_+) \omega$. 
The general solutions, $\Phi_{S1}^{(0)}$ and $\Phi_{S2}^{(0)}$, are given for the near-extremal $\sigma \neq 0$ case:
\bena
\Phi_{S1}^{(0)}
 &=& C_1\cdot {}_2F_1\left( - \nu + \frac{1}{2}, \nu+\frac{1}{2}, 1+ 2i \frac{\omega}{\sigma}; 1 + \frac{x}{\sigma} \right) 
 \non \\
  &{}&  
     + C_2 \cdot (x+ \sigma)^{-2i \omega/\sigma} 
             {}_2F_1\left( - \nu + \frac{1}{2}- 2i \frac{\omega}{\sigma}, \nu+\frac{1}{2}- 2i \frac{\omega}{\sigma}, 1- 2i \frac{\omega}{\sigma}; 1 + \frac{x}{\sigma}  \right) \,, 
\\
\Phi_{S2}^{(0)}
 &=& C_1\cdot {}_2F_1\left( - \nu + \frac{3}{2}, \nu+\frac{3}{2}, 2 + 2i \frac{\omega}{\sigma}; 1 + \frac{x}{\sigma} \right) 
 \non \\
  &{}&  
     + C_2 \cdot (x+ \sigma)^{-1-2i \omega/\sigma} 
             {}_2F_1\left( - \nu + \frac{1}{2}- 2i \frac{\omega}{\sigma}, \nu+\frac{1}{2}- 2i \frac{\omega}{\sigma}, -2i \frac{\omega}{\sigma}; 1 + \frac{x}{\sigma}  \right) \,,              
\eena
where 
\bena
  \nu := \sqrt{k_S^2 +\mu^2r_+^2 + \frac{1}{4}} \,.  
\eena
As for the extremal $\sigma=0$ case:
\bena
 \Phi_{S1}^{(0)}
 &=&  \frac{1}{\sqrt{x}}e^{-i\omega/x}
   \left[
           C_1\cdot I_\nu \left( i \omega/x \right) +   C_2\cdot K_\nu \left( i \omega/x \right) 
   \right] \,,
\eena
and
\bena
\Phi_{S2}^{(0)}
 &=&  
           C_1\cdot {x}^{-5/2}e^{-i\omega/x}\cdot
                          \left\{ 
                                   \omega I_{\nu +1} \left(-i \omega/x \right) 
                                   + i \left[(\nu + 1/2) x - i\omega  \right]  I_{\nu} \left(- i \omega/x \right)
                         \right\} 
\non \\
  &{+}&  C_2\cdot {x}^{-5/2}e^{-i\omega/x}\cdot
                         \left\{ 
                                  - \omega K_{\nu +1} \left(-i \omega/x \right) 
                                   + i \left[(\nu + 1/2) x - i\omega  \right]  K_{\nu} \left(-i \omega/x \right)
                         \right\} \,. 
\eena
With the choice of boundary conditions of interest, one can 
construct the Green's functions $G_{\alpha 1}^{(0)} = L_{\alpha 1}^{(0)}{}^{-1}$ and $G_{\beta 2}^{(0)} = L_{\beta 2}^{(0)}{}^{-1}$. 
Then, one can obtain the $n$-th order solutions, formally expressed as
\bena
\Phi_{S1}^{(n)} &=& -G_{\alpha 1}^{(0)}  \sum^{n}_{m=1} 
                             \left\{ 
                                      L^{(m)}_{\alpha 1}\Phi_{S1}^{(n-m)}+L^{(m)}_{\alpha 2}\Phi_{S2}^{(n-m)}+L^{(m)}_{\alpha 3}\Phi_{S3}^{(n-m)}
                             \right\} \,,
\\
\Phi_{S2}^{(n)} &=& -G_{\beta 2}^{(0)}  \sum^{n}_{m=1} 
                             \left\{ 
                                      L^{(m)}_{\beta 2}\Phi_{S2}^{(n-m)}+L^{(m)}_{\beta 3}\Phi_{S3}^{(n-m)}
                             \right\} \,. 
\eena

\subsection{Proca equations in general (near-)extremal black holes}

In this subsection, we provide the expansion of the Proca equation in more generic, 
extremal and near-extremal black holes in four-dimensions. Our metric ansatz is given by 
eq.~(\ref{def:metric:lambda}). We expand the general metric functions, 
$F(\lambda x)=x(x+\sigma)g(\lambda x)$ and $R(\lambda x)$, as
\ben
g=\sum_{n=0}^{\infty}\lambda^{n}g^{(n)} \,, \quad R=\sum_{n=0}^{\infty}\lambda^{n}R^{(n)} \,, 
\label{exp:g:R} 
\een
where $g^{(0)}, \: R^{(0)}$ are assumed to be some positive constants 
and $g^{(n)}, \:R^{(n)} \: (n \geqslant 1)$ can be any regular (except at a singularity) functions of $x$.  
%
For later use, we define the following quantities: 
\bena
\Delta^{(n)}&:=& \sum_{m=0}^{n}\sum_{l=0}^{m}R^{(n-m)}R^{(m-l)}g^{(l)} \,, 
\label{eq:Delta}
\\
\Delta^{(n)}_{g}&:=& \sum_{m=0}^{n}\sum_{l=0}^{m}R^{(n-m)}R^{(m-l)}(\partial_{x}g^{(l)}) \,, 
\label{eq:Delta_g}
\\
\Delta^{(n)}_{gg}&:=& \sum_{m=0}^{n}\sum_{l=0}^{m}R^{(n-m)}R^{(m-l)}(\partial_{x}^{2}g^{(l)}) \,, 
\label{eq:Delta_gg}
\\
\Delta^{(n)}_{R}&:=& \sum_{m=0}^{n}\sum_{l=0}^{m}R^{(n-m)}(\partial_{x}R^{(m-l)})g^{(l)} \,, 
\label{eq:Delta_R}
\\
\Delta^{(n)}_2 &:=& \sum_{m=0}^{n}R^{(n-m)}R^{(m)} \,, 
\label{eq:delta}
\\
\Delta^{(n)}_{2R}&:=& \sum_{m=0}^{n}R^{(n-m)}(\partial_{x}R^{(m)}) \,. 
\label{eq:delta_R}
\eena

\subsubsection{Vector-type component of the Proca equation}
The master equation for the vector-type component (\ref{eq:master:vect:m}) becomes in the present case 
\ben
\left[
       R^{2}F\partial_{x}^{2}+R^{2}(\partial_{x}F)\partial_{x} +2R^{2}\partial_{v}\partial_{x}
       -\left(
                k^{2}_{V}+1+\mu^{2} r_{+}^{2} R^{2}
         \right)
\right]\phi^V=0 \,.
\label{eq:general_Proca_vector}
\een
As we have done in the previous subsection, we expand this equation with respect to $\lambda$. 
%
%
Let us introduce the operator: 
\bena
\mathcal{L}^{(n)}_{V}
   &:=&  \Delta^{(n)} x(x+\sigma)\partial_{x}^{2}
                                         + \left\{ 
                                                     \Delta^{(n)}_{g} x (x+\sigma) + \Delta^{(n)}(2x+\sigma)+2\Delta_2^{(n)}\partial_{v}
                                            \right\} \partial_{x}
\non
\\
   &{}& \quad 
     - \delta_{n0}(k^{2}_{V}+1) - \Delta_2^{(n)}\mu^{2}r_{+}^{2} \,.
\label{eq:general_Proca_vector_op}
\eena
In terms of this operator, we obtain  
\ben
\sum^{\infty}_{n=0}\lambda^{n}\left[\sum^{n}_{m=0}\mathcal{L}^{(m)}_{V}\Phi_V^{(n-m)}\right]=0
\,.
\label{eq:general_Proca_vec_expand}
\een
Thus, at $n$-th order, we have 
\ben
\sum^{n}_{m=0}\mathcal{L}^{(m)}_{V}\Phi_V^{(n-m)}=0 \,.
\label{eq:Proca_vec_expand2}
\een
%
%
Thus, we obtain the set of master equations for the vector-type component 
 with the operators $L^{(n)}_V$'s in eqs.~(\ref{eq:v:0}) -- (\ref{eq:v:n}) replaced with 
${\cal L}^{(n)}$'s defined in eq.~(\ref{eq:general_Proca_vector_op}). 
Once the leading master variable $\Phi_V^{(0)}$ is obtained, 
one can successively obtain the solution at any order $\Phi_V^{(n)}$. 

\subsubsection{Scalar-type component of the Proca equation}

In our present general four-dimensional background, the scalar-type component 
of the Proca equation, (\ref{eq:Proca_scalar1}) and (\ref{eq:Proca_scalar2}), 
are rewritten as  
\bena
&&
    \left[ R^{2}F\partial_{x}^{2}+R^{2}(\partial_{x}F)\partial_{x}
           + 2R^{2}\partial_{v}\partial_{x} + 2R(\partial_{x}R)(F\partial_{x}+\partial_{v})
           - \left(
                     k^{2}_{S}+\mu^{2}r_{+}^{2} R^{2} 
             \right)
    \right]A
\non 
\\
&{}& \hspace{8cm}
+2R(\partial_{x}R)(FB_{x}+B_{v})=0 \,,
%
\label{eq:general_Proca_scalar1_ver2}
\eena
\bena
&& \left[
              R^{2}F\partial_{x}^{2}+2R^{2}(\partial_{x}F)\partial_{x}
            +R^{2}(\partial_{x}^{2}F)\partial_{x}
+2R^{2}\partial_{v}\partial_{x}+2R^{2}(\partial_{x}R)\partial_{v}-\left(k^{2}_{S}+\mu^{2} r_{+}^{2}R^2\right)
     \right]B_{x}
\non \\
&{}& \hspace{9cm}
  -2R(\partial_{x}R)\partial_{x}B_{v}=0 \,,
%
\label{eq:general_Proca_scalar2_ver2-1}
\eena
\bena
&& 
\left[
          R^{2}F\partial_{x}^{2}+2R^{2}\partial_{v}\partial_{x}
        +2R(\partial_{x}R)F\partial_{x} - \left(k^{2}_{S}+\mu^{2}r_{+}^{2} R^2\right) 
\right]B_{v}
\non \\
&{}& \hspace{6cm}
  + \left[
            R^{2}(\partial_{x}F)
             -2R(\partial_{x}R) F
    \right] \partial_{v}B_{x}=0 \,.
%
\label{eq:general_Proca_scalar2_ver2-2}
\eena
As in the vector-type case, we expand these equations by $\lambda$. 
The three variables $A,\: B_x,\: B_v$ are expanded as 
(\ref{eq:A_expand}), (\ref{eq:Bx_expand}), and (\ref{eq:Bv_expand}).
%
We define the set of operators:  
\bena 
\mathcal{L}^{(n)}_{\alpha 1} 
   &:=& \Delta^{(n)} x(x+\sigma)\partial_{x}^{2}
          +\left\{ 
                     (\Delta^{(n)}_{g} + 2\Delta^{(n)}_{R})x(x+\sigma)+\Delta^{(n)}(2x+\sigma)+2\Delta_2^{(n)}\partial_{v}
            \right\} \partial_{x}
\non \\
&& \,
 +2\Delta^{(n)}_{2R}\partial_{v}-\delta_{n0}k^{2}_{S} - \Delta_2^{(n)}\mu^{2}r_{+}^{2} \,,
\label{eq:general_aA_op}
\\
&& \non \\  
\mathcal{L}^{(n)}_{\alpha 2} &:=& 2\Delta^{(n)}_{R}x(x+\sigma) \,,
\label{eq:general_ax_op}
\\
&& \non \\ 
\mathcal{L}^{(n)}_{\alpha 3} &:=& 2 \Delta^{(n)}_{2R} \,,
\label{eq:general_av_op}
\\
&& \non \\ 
\mathcal{L}^{(n)}_{\beta 2} &:=& \Delta^{(n)} x(x+\sigma)\partial_{x}^{2}+2\{\Delta^{(n)}_{g}x(x+\sigma)+\Delta^{(n)}(2x+\sigma)+\Delta_2^{(n)}\partial_{v}\}\partial_{x}
\non 
\\
&{}& \, 
 +\Delta^{(n)}_{gg}x(x+\sigma)+2\Delta^{(n)}_{g}(2x+\sigma)+2\Delta^{(n)}+2\Delta^{(n)}_{2R}\partial_{v}-\delta_{n0}k^{2}_{S}
-\Delta_2^{(n)}\mu^{2}r_{+}^{2} \,, 
\label{eq:general_bx_op}
\\
&& \non \\ 
\mathcal{L}^{(n)}_{\beta 3} &:=& 2\Delta^{(n)}_{2R}\partial_{x} \,,
\label{eq:general_bv_op}
\\
&& \non \\ 
\mathcal{L}^{(n)}_{\gamma 2} &:=& \{(\Delta^{(n)}_{g}-2\Delta^{(n)})x(x+\sigma)+\Delta^{(n)}(2x+\sigma)\}\partial_{v} \,,
\label{eq:general_cx_op}
\\ 
&& \non \\ 
\mathcal{L}^{(n)}_{\gamma 3} &:=& \Delta^{(n)} x(x+\sigma)\partial_{x}^{2}+2\{\Delta^{(n)}x(x+\sigma)
+\Delta_2^{(n)}\partial_{v}\}\partial_{x}-\delta_{n0}k^{2}_{S} - \Delta_2^{(n)}\mu^{2}r_{+}^{2} \,.
\label{eq:general_cv_op}
\eena
The analysis essentially parallels that of the Reissner-Nordstrom case. 
From eqs.~(\ref{eq:general_Proca_scalar1_ver2}), 
(\ref{eq:general_Proca_scalar2_ver2-1}), and 
(\ref{eq:general_Proca_scalar2_ver2-2}), we have at $n$-th order, the same formulas as (\ref{eq:Proca_scalar1_ver4}), (\ref{eq:Proca_scalar2_ver4-1}), and (\ref{eq:Proca_scalar2_ver4-2})
with $L^{(n)}$'s replaced with ${\cal L}^{(n)}$'s given above: 
%
%
\ben
\sum^{n}_{m=0}\left\{\mathcal{L}^{(m)}_{\alpha 1}\Phi_{S1}^{(n-m)}+\mathcal{L}^{(m)}_{\alpha 2}\Phi_{S2}^{(n-m)}+\mathcal{L}^{(m)}_{\alpha 3}\Phi_{S3}^{(n-m)}\right\}=0
\label{eq:general_Proca_scalar1_ver4}
\een
\ben
\sum^{n}_{m=0}\left\{\mathcal{L}^{(m)}_{\beta 2}\Phi_{S2}^{(n-m)}+\mathcal{L}^{(m)}_{\beta 3}\Phi_{S3}^{(n-m)}\right\}=0
\,,
\label{eq:general_Proca_scalar2_ver4-1}
\een
\ben
\sum^{n}_{m=0}\left\{\mathcal{L}^{(m)}_{\gamma 2}\Phi_{S2}^{(n-m)}+\mathcal{L}^{(m)}_{\gamma 3}\Phi_{S3}^{(n-m)}\right\}=0
\,.
\label{eq:general_Proca_scalar2_ver4-2}
\een
More explicitly, 
\begin{itemize}
\item[(i)] At the leading-order $\lambda=0$, the background is the corresponding near-horizon geometry, and we find 
\ben
\mathcal{L}^{(0)}_{\alpha 2}=\mathcal{L}^{(0)}_{\alpha 3}=\mathcal{L}^{(0)}_{\beta 3}=0 \,.
\een
Therefore we have 
\ben
\left(
\begin{tabular}{ccc}
$\mathcal{L}^{(0)}_{\alpha 1}$&$0$&$0$\\
$0$&$\mathcal{L}^{(0)}_{\beta 2}$&$0$\\
$0$&$\mathcal{L}^{(0)}_{\gamma 2}$&$\mathcal{L}^{(0)}_{\gamma 3}$\\
\end{tabular}
\right)
\left(
\begin{tabular}{c}
$\Phi_{S1}^{(0)}$\\
$\Phi_{S2}^{(0)}$\\
$\Phi_{S3}^{(0)}$\\
\end{tabular}
\right)
=
\left(
\begin{tabular}{c}
$0$\\
$0$\\
$0$\\
\end{tabular}
\right) \,. 
\label{eq:general_Proca_scalar_0}
\een
This is the set of leading-order, decoupled master equations for two master 
variables, $\Phi_{S1}^{(0)}, \: \Phi_{S2}^{(0)}$. The remaining variable $\Phi_{S3}^{(0)}$ can be 
expressed in terms of $\Phi_{S2}^{(0)}$. 

\item[(ii)] In general, at $n$-th order, we have 
\ben
\left(
\begin{tabular}{ccc}
$\mathcal{L}^{(0)}_{\alpha 1}$&$0$&$0$\\
$0$&$\mathcal{L}^{(0)}_{\beta 2}$&$0$\\
$0$&$\mathcal{L}^{(0)}_{\gamma 2}$&$\mathcal{L}^{(0)}_{\gamma 3}$\\
\end{tabular}
\right)
\left(
\begin{tabular}{c}
$\Phi_{S1}^{(n)}$\\
$\Phi_{S2}^{(n)}$\\
$\Phi_{S3}^{(n)}$\\
\end{tabular}
\right)
=
-
\sum_{m=1}^{n}
\left[
\left(
\begin{tabular}{ccc}
$\mathcal{L}^{(m)}_{\alpha 1}$&$\mathcal{L}^{(m)}_{\alpha 2}$&$\mathcal{L}^{(m)}_{\alpha 3}$\\
$0$&$\mathcal{L}^{(m)}_{\beta 2}$&$\mathcal{L}^{(m)}_{\beta 3}$\\
$0$&$\mathcal{L}^{(m)}_{\gamma 2}$&$\mathcal{L}^{(m)}_{\gamma 3}$\\
\end{tabular}
\right)
\left(
\begin{tabular}{c}
$\Phi_{S1}^{(n-m)}$\\
$\Phi_{S2}^{(n-m)}$\\
$\Phi_{S3}^{(n-m)}$\\
\end{tabular}
\right)\right] \,. 
\label{eq:general_Proca_scalar_n}
\een
\end{itemize} 
This is the set of master equations for the scalar-type components of the Proca 
equation in the general static black hole background, in which we have, at any order, 
the formulas similar to those obtained in the Reissner-Nordstrom case 
with the operators $L^{(n)}$'s replaced with ${\cal L}^{(n)}$'s. Namely, we have two 
mutually decoupled master equations with source terms that consist only of 
the lower-order variables, and therefore, in principle, once having obtained 
the solutions to the leading-order 
homogeneous master equations, one can successively obtain the solutions 
to any order of the scalar-type components of the Proca equation. 
These formulas are our main results.

\section{Summary and Discussion}
\label{sect:Summary}
We have developed a new perturbation method to solve the Proca equation in static 
extremal and near-extremal black hole spacetimes, providing for the first time 
a set of mutually decoupled wave equations for massive vector field at each order of 
perturbations. Our formulas can be a useful tool to 
analytically study the behavior of massive vector fields around 
(near-)extremal black holes.
We have first considered the background metric which takes the warped product form 
of an $m$-dimensional arbitrary spacetime ${\cal N}^m$ and an $n$-dimensional Einstein space 
${\cal K}^n$, which essentially describes the horizon cross-section manifold. 
We have classified the massive vector field variables into the vector-type 
and scalar-type components according to their behavior on the Einstein space ${\cal K}^n$. 
Then, by introducing the scalar and vector harmonics on ${\cal K}^n$, 
we have separated the field variables and reduced the Proca equation to the set of wave 
equations on the generic spacetime ${\cal N}^m$. 
At this stage, the Proca equation for the scalar-type and vector-type variables 
are decoupled each other. Furthermore, for the vector-type components 
we have derived the single master equation (\ref{eq:master:vect:m}) 
for the master variable $\Phi^V$ on the generic spacetime ${\cal N}^m$. 
On the other hand, at this stage, for the scalar-type components, we have obtained 
the set of coupled wave equations, (\ref{eq:Ba}) and (\ref{eq:A}), 
for the variables $(B_a, A)$ on the generic background ${\cal N}^m$. Note however that for the massless case, 
i.e., Maxwell field, by exploiting the recovered gauge freedom, we have also been able to derive 
the single master equation, (\ref{eq:master:maxwell}), for the scalar-type components of the Maxwell field 
on the generic warped product background ${\cal N}^2 \times {\cal K}^n$.    

\medskip 

In order to obtain a set of decoupled wave equations for the scalar-type components of massive vector field, 
we have restricted our attention to the extremal and near-extremal static black hole background 
with $m=2$, in which ${\cal N}^2$ is spanned by the advanced time and 
radial coordinates $y^a=(v,x)$. Such a (near-) extremal black hole 
admits the near-horizon limit $\lambda \rightarrow 0$, which is known to possess 
enhanced symmetry. 
We have viewed the scaling parameter $\lambda$ as a small perturbation parameter,
and expanded the massive vector field variables as well as 
the background metric components about $\lambda$, with the leading-order 
geometry being the corresponding near-horizon geometry. Then, we have derived the set of 
wave equations for the massive vector field perturbations at each order of $\lambda$. 
At the leading-order, we have found that thanks to the enhanced symmetry of 
the near-horizon geometry, the scalar-type components of the Proca equation 
reduce to the two mutually decoupled homogeneous master equations 
for $(\Phi_{S1}^{(0)}, \: \Phi_{S2}^{(0)})$, and that the remaining component $\Phi_{S3}^{(0)}$ can be 
determined by these two master variables.  
Then, we have also found that at any higher (say, $n$-th) order, 
the scalar-type components of the Proca equation always reduce to the two mutually 
decoupled inhomogeneous wave equations for the two master scalars 
$(\Phi_{S1}^{(n)}, \: \Phi_{S2}^{(n)})$ with source terms that consist only of 
the lower-order variables, and the remaining component at the $n$-th order $\Phi_{S3}^{(n)}$ 
can be determined by the master variables at the same and lower-order. 
Therefore, once we solve the leading order homogeneous master equations on the near-horizon 
geometry, we can, in principle, solve successively the set of inhomogeneous wave equations at any order. 
With the vector-type and scalar-type components all together, at each order the triplet $(\Phi_V^{(n)}, \: \Phi_{S1}^{(n)}, \:\Phi_{S2}^{(n)})$ describes the three independent dynamical degrees of freedom for massive vector field.   
We have provided the general solutions, $(\Phi_V^{(0)}, \: \Phi_{S1}^{(0)}, \:\Phi_{S2}^{(0)})$, 
to the leading order master equations for the extremal and near-extremal 
Reissner-Nordstrom black hole case.

\medskip 

In this paper, we have focused on the static background case.   
In astrophysical context, extremal or near-extremal rotating black holes are 
more relevant. In the rotating case, it is far from obvious {\it a priori} whether it is possible 
to separate field variables of interest. Recently it has been shown by developing the new 
ansatz~\cite{Lunin:2017} that massive vector field equations 
can be separable in Kerr-NUT-(A)dS spacetimes~\cite{Frolov:Krtous:Kubiznak:Santos:2018}. 
Since the class of spacetimes dealt with in~\cite{Frolov:Krtous:Kubiznak:Santos:2018} 
does not contain {\em static extremal} black holes considered in the present paper, one 
cannot immediately compare the result of~\cite{Frolov:Krtous:Kubiznak:Santos:2018} and that 
of the present paper. However, as explicitly stated in~\cite{Frolov:Krtous:Kubiznak:Santos:2018} that 
their solutions describe (in even $D$-dimensions) $D-2$ real modes, but one polarization is 
missing. It has been shown, more concretely, by considering the four-dimensional Kerr black hole background 
and numerically computing quasi-normal modes~\cite{Frolov:Krtous:Kubiznak:Santos:2018} 
that their separation ansatz allows one to derive a decoupled equation which correctly describe, at least, two of the three physical polarizations of the massive vector field, but how to obtain the remaining polarization within their ansatz remains open. 
In contrast, although in this paper we have restricted our attention only to a class of static and (near-) extremal black hole backgrounds, 
we have successfully been able to obtain, at each order of perturbations, three decoupled equations for all the three polarizations of massive vector perturbations. Therefore, at the present stage, it is fair to say that our method developed in this paper and the ansatz of \cite{Frolov:Krtous:Kubiznak:Santos:2018} are regarded as complementary methods. 
%
Thus, the remaining open issue is whether (and if possible how) one can 
derive a set of decoupled wave equations for all the three (in four-dimensions) physical polarizations 
in the rotating black hole case. 
It is of considerable interest to generalize to the maximally rotating Kerr black hole 
case our method of expanding both the field variables and background geometry with respect to 
the near-horizon scaling parameter and exploiting the enhanced symmetry of the near-horizon geometry 
to obtain the leading order solutions. It would also be interesting--even for the static case--to consider 
a generalization of the present method to systems of, e.g., Maxwell theory with the Chern-Simons term. 

\bigskip 

\noindent
{\bf Acknowledgements:} 
We would like to thank Tomoki Minamigawa and Takashi Okumura for discussions. 
The work of A.I. was supported in part by JSPS KAKENHI Grants 
No.~15K05092 and No.~26400280.

\appendix
\section{Massive charged scalar field on (near-)extremal Reissner-Nordstrom black holes}

In this appendix, we apply our perturbation method to a charged massive scalar field  
in the near-extremal Reissner-Nordstrom black hole background. 
Let us consider the background metric,  (\ref{eq:(vx)metric}) and (\ref{eq:RN_F_R}). 
On this background, we consider the  massive charged scalar field that obeys the Klein-Gordon equation: 
\ben
(\mathcal{D}_{\mu}\mathcal{D}^{\mu}-\mu^{2})\psi=0 \,, 
\label{eq:KG}
\een
\ben
\mathcal{D}_{\mu} := \nabla_{\mu}-iq{\cal A}_{\mu} \,, 
\een
where $q$ denotes the coupling constant to the background gauge-field ${\cal A}_\mu$, which 
can take, under a certain gauge,  the following form:  
\ben
{\cal A}_\mu dx^\mu =\frac{Qx}{1+\lambda x}dv \,.
\een 
The Klein-Gordon equation above is written as
\ben
\left[F \partial_{x}^{2} + (\partial_{x}F)\partial_{x} + 2 \partial_{v}\partial_{x}-2iqQ\left(\frac{x}{1+\lambda x}\right)\partial_{x}
       - \left\{
                   \frac{k_S^2+iqQ}{(1+\lambda x)^2} + \mu^2 r_+^2
         \right\}
       \right]\psi=0 \,.
\label{eq:general_KGver2_2}
\een
We expand the scalar field as
\ben
\psi=\sum_{n=0}^{\infty} \lambda^{n}\psi^{(n)} \,, 
\label{eq:psi_expand}
\een
and also the above equation (\ref{eq:general_KGver2_2}) with respect to $\lambda$. 
Then, defining the operators, 
\bena
 L^{(n)}_{\psi} &:=& (-1)^{n}
                      \Big[ 
                             (n+1) x^{n+1}(x+\sigma)\partial_{x}^{2}+(n+1)x^{n} \left\{(n+2)x+(n+1)\sigma \right\} \partial_{x} 
\non \\
&{}& \qquad \quad 
                           + 2\delta_{n0}\partial_{v}\partial_{x}-2iqQx^{n+1}\partial_{x}-(n+1)(k^{2}_{S}+iqQ)x^{n}-\delta_{n0}\mu^{2}r_{+}^{2}
                      \Big] \,,  
\label{eq:psi_op}
\eena
we can write eq.~(\ref{eq:general_KGver2_2}) as 
\ben
\sum^{\infty}_{n=0}\lambda^{n}\left[\sum^{n}_{m=0} {L}^{(m)}_{\psi}\psi^{(n-m)}\right]=0 \,,
\label{eq:general_KG_expand}
\een
and therefore obtain the formulas:  
\ben
\sum^{n}_{m=0} {L}^{(m)}_{\psi}\psi^{(n-m)}=0 \,.
\label{eq:Proca_vec_expand2}
\een
More explicitly, we have 
\bena
{L}^{(0)}_{\psi}\psi^{(0)}&=&0 \,, 
\label{eq:leading:psi}
\\
 {L}^{(0)}_{\psi}\psi^{(1)}&=& - {L}^{(1)}_{\psi}\psi^{(0)} \,, 
\\
 {L}^{(0)}_{\psi}\psi^{(2)}&=& - {L}^{(1)}_{\psi}\psi^{(1)} - {L}^{(2)}_{\psi}\psi^{(0)} \,, 
\\
&\vdots&\non
\\
 {L}^{(0)}_{\psi}\psi^{(n)}&=& - \sum^{n}_{m=1} {L}^{(m)}_{\psi}\psi^{(n-m)} \,.
\eena 
Thus, as in the Proca equation case, once we obtain the leading-order solution, $\psi^{(0)}$, 
we can successively solve the Klein-Gordon equation (\ref{eq:Proca_vec_expand2}), and 
obtain solution $\psi^{(n)}$ at any higher orders. 

\medskip 

We also provide here the general solution to the leading-order equation, $L_\psi^{(0)} \psi^{(0)}=0$. 
With the ansatz of time-dependency $\psi \propto e^{-i\omega v}$, the time-derivative $\partial_v$ is replaced with 
$-i\omega$ and the leading operator becomes the second-order ordinary differential operator:  
\ben
 L_\psi^{(0)} = x(x+\sigma) \frac{d^2}{dx^2} + 2\left[ (1-iqQ)x + \sigma/2-i\omega \right] \frac{d}{dx} - (k_S^2+ \mu^2r_+^2+iqQ) \,.
\een
As the time-coordinate $v$ is scale-transformed as $v \rightarrow (r_+ /\lambda) v $ in eq.~(\ref{def:scaling}), the frequency $\omega$ is also scale-transformed as $\omega \rightarrow (\lambda/r_+) \omega$. 
The general solution is then, for the near-extremal $\sigma \neq 0$ case:
\bena
\psi^{(0)} \!
 &=& \!C_1\cdot {}_2F_1\left( - \nu + \frac{1}{2}-iqQ, \nu+\frac{1}{2}-iqQ, 1+2i \kappa; 1 + \frac{x}{\sigma} \right) 
 \non \\
  &+& \! 
     C_2 \cdot (x+ \sigma)^{-2i \kappa} 
       {}_2F_1\left( - \nu + \frac{1}{2}+iqQ-2i \frac{\omega}{\sigma}, \nu+\frac{1}{2} + iqQ - 2i \frac{\omega}{\sigma}, 1-2i \kappa; 1 + \frac{x}{\sigma}  \right) \,, 
\eena
where 
\bena
  \nu := \sqrt{k_S^2+\mu^2r_+^2 -q^2Q^2 + \frac{1}{4}} \,,  \quad 
  \kappa := \frac{\omega}{\sigma} -qQ \,.
\eena
As for the extremal $\sigma=0$ case:
\bena
 \psi^{(0)}
 &=&  
    x^{iqQ} e^{-i\omega/x}
   \left[
           C_1\cdot M_{-iqQ,\nu} \left(2 i \omega/x \right) +   C_2\cdot W_{-iqQ,\nu} \left(2 i \omega/x \right) 
   \right] \,,
\eena
where $M_{-iqQ,\nu}, \:W_{-iqQ,\nu}$ denote the Whittaker functions. By constructing the Green function $G_\psi^{(0)}= L_\psi^{(0)}{}^{-1}$, 
one can obtain the $n$-th order solution as 
\ben
 \psi^{(n)} = - G_\psi^{(0)} \sum^{n}_{m=1} {L}^{(m)}_{\psi}\psi^{(n-m)} \,. 
\een


\begin{thebibliography}{99}

\bibitem{Arvanitaki:2009fg}
  A.~Arvanitaki, S.~Dimopoulos, S.~Dubovsky, N.~Kaloper, J.~March-Russell,
  Phys.\ Rev.\  D{\bf 81}, 123530 (2010).
  [arXiv:0905.4720 [hep-th]].
  
\bibitem{Acharya:2015zfk}
B.S.~Acharya, C.~Pongkitivanichkul,
JHEP 1604 (2016) 009

\bibitem{Goodsell:2009xc}
M.~Goodsell, J.~Jaeckel, J.~Redondo, A.~Ringwald 
JHEP 0911 (2009) 027

\bibitem{Arvanitaki:2010sy}
A.~Arvanitaki, S.~Dubovsky,
Phys. Rev. D{\bf 83} 044026  (2011)
 
\bibitem{Press:1972zz}
  W.~H.~Press, S.~A.~Teukolsky,
  Nature {\bf 238}, 211-212 (1972).
 

\bibitem{Cardoso:2004nk}
 V. Cardoso, O. J. C. Dias, J. P. S. Lemos, and S. Yoshida, 
 Phys. Rev. D{\bf 70} (2004) 044039,
arXiv:hep-th/0404096 [hep-th]. [Erratum: Phys. Rev.D70,049903(2004)].
 
 
\bibitem{Herdeiro:DR2013}
C. A. R. Herdeiro, J. C. Degollado, and H. F. Rnarsson, 
Phys. Rev. D{\bf 88} (2013) 063003,
arXiv:1305.5513 [gr-qc]. 
 
\bibitem{Hod:2013a}
S. Hod, 
Phys. Rev. D{\bf 88} (2013) 064055, 
arXiv:1310.6101 [gr-qc].

\bibitem{Degollado:Herdeiro:2014}
J. C. Degollado and C. A. R. Herdeiro, 
Phys. Rev. D89 (2014) no. 6, 063005,
arXiv:1312.4579 [gr-qc].

\bibitem{LiZhao:2015}
R. Li and J. Zhao, 
Phys. Lett. B740 (2015) 317–321, arXiv:1412.1527 [gr-qc].
 
 
 
\bibitem{Hawking:Reall:2000}
S. W. Hawking and H. S. Reall,
Phys. Rev. D61 (2000) 024014, 
arXiv:hep-th/9908109 [hep-th].

\bibitem{Cardoso:2004hs}
  V.~Cardoso, O.~J.~C.~Dias,
  Phys.\ Rev.\  {\bf D70}, 084011 (2004).
  [hep-th/0405006].


\bibitem{Cardoso:2006wa}
  V.~Cardoso, O.~J.~C.~Dias, S.~Yoshida,
  Phys.\ Rev.\  {\bf D74}, 044008 (2006).
  [hep-th/0607162].

\bibitem{Uchikata:2009zz}
  N.~Uchikata, S.~Yoshida, T.~Futamase,
  Phys.\ Rev.\  {\bf D80}, 084020 (2009).


\bibitem{Kodama:2009rq}
  H.~Kodama, R.~A.~Konoplya, A.~Zhidenko,
  Phys.\ Rev.\  {\bf D79}, 044003 (2009).
  [arXiv:0812.0445 [hep-th]].
  
\bibitem{Cardoso:Dias:Hartnett:Lehner:Santos:2014}
V. Cardoso, O. J. C. Dias, G. S. Hartnett, L. Lehner, and J. E. Santos, 
JHEP 04 (2014) 183,
arXiv:1312.5323 [hep-th]. 

\bibitem{Green:Hollands:Ishibashi:Wald:2016}
S.R.~Green, S.~Hollands, A.~Ishibashi, R.M.~Wald, 
Class. Quant. Grav. {\bf 33} (2016) no.12, 125022

 
 
 
\bibitem{Damour:1976kh}
  T.~Damour, N.~Deruelle, R.~Ruffini,
  Lett.\ Nuovo Cim.\  {\bf 15}, 257-262 (1976).
  
\bibitem{Zouros:1979iw}
  T.~J.~M.~Zouros, D.~M.~Eardley,
  Annals Phys.\  {\bf 118}, 139-155 (1979).

\bibitem{Detweiler:1980uk}
  S.~L.~Detweiler,
  Phys.\ Rev.\  {\bf D22}, 2323-2326 (1980). 

\bibitem{Dolan:2007mj}
  S.~R.~Dolan,
  Phys.\ Rev.\  {\bf D76}, 084001 (2007).
  [arXiv:0705.2880 [gr-qc]].


\bibitem{Rosa:2009ei}
  J.~G.~Rosa,
  JHEP {\bf 1006}, 015 (2010).
  [arXiv:0912.1780 [hep-th]].


\bibitem{Hod:2011}
S.~Hod, 
Phys. Rev. D{\bf 84} (2011) 044046


\bibitem{Pani:2012vp}
P. Pani, V. Cardoso, L. Gualtieri, E. Berti, and A. Ishibashi, 
Phys. Rev. Lett. 109 (2012) 131102, 
arXiv:1209.0465 [gr-qc].


\bibitem{Pani:2012bp}
P. Pani, V. Cardoso, L. Gualtieri, E. Berti, and A. Ishibashi, 
Phys.Rev. D86 (2012) 104017, arXiv:1209.0773 [gr-qc].


\bibitem{Witek:2012tr}
H. Witek, V. Cardoso, A. Ishibashi, and U. Sperhake, 
Phys. Rev. D{\bf 87} (2013) 043513, 
arXiv:1212.0551 [gr-qc].

\bibitem{Brito:CP:2013}
R. Brito, V. Cardoso, and P. Pani, 
Phys.Rev. D88 (2013) 023514, 
arXiv:1304.6725 [gr-qc].



\bibitem{Yoshino:Kodama:2015}
H. Yoshino and H. Kodama, 
arXiv:1505.00714 [gr-qc]
 

\bibitem{Hod:2016}
S.~Hod, 
Phys.Rev. D{\bf 94} (2016) 044036

\bibitem{Ishibashi:Pani:Gualtieri:Cardoso:2015} 
A.~Ishibashi, P.~Pani, L.~Gualtieri, V.~Cardoso, 
JHEP 1509 (2015) 209




\bibitem{Cardoso:2018tly}
V.~Cardoso, Ó.J.C. Dias, G.S. Hartnett, M. Middleton, P. Pani, J.E. Santos,
JCAP 1803 (2018) no.03, 043

\bibitem{Herdeiro:Radu2014}
C.A.R. Herdeiro and E. Radu, 
Phys. Rev. Lett. {\bf 119} (2017) 261101 
[arXiv:1706.06597]

\bibitem{Herdeiro:Radu2017}
C.A.R. Herdeiro and E. Radu, 
Phys. Rev. Lett. {\bf 112} (2014) 221101 [arXiv:1403.2757]

\bibitem{Ganchev:Santos:2017}
B. Ganchev and J.E. Santos, 
arXiv:1711.08464

\bibitem{Degollado:Herdeiro:Radu:18} 
J.C. Degollado, C.A.R. Herdeiro and E. Radu, 
arXiv:1802.07266

\bibitem{Lunin:2017}
O.~Lunin, 
JHEP 12 (2017) 138


\bibitem{Konoplya:2006}
R.A.~Konoplya, 
Phys. Rev. D{\bf 73}, 024009 (2006). 


\bibitem{Konoplya:Zhidenko:Molina:2007}
R.A.~Konoplya, A. Zhidenko, C.~Molina,
Phys. Rev. D{\bf 75}, 084004 (2007). 


\bibitem{Rosa:Dolan:2012}
J.G.~Rosa and S.R.~Dolan,
Phys. Rev. D{\bf 85}, 044043 (2012). 


\bibitem{Zilhao:Witek:Cardoso:2015}
M.~Zilhao, H.~Witek, V.~Cardoso, 
 Class.\ Quant.\ Grav.\  {\bf 32},  234003 (2015)


\bibitem{East:Pretorius:2017}
W.E.~East and F.~Pretorius, 
Phys. Rev. Lett. {\bf 119} (2017) 041101


\bibitem{Kojima:92}
Y. Kojima, Phys. Rev. D{\bf 46}, 4289 (1992).

\bibitem{Kojima:apj:93}
Y. Kojima, Astrophys. J. {\bf 414}, 247 (1993)

\bibitem{Kojima:ptp:93}
Y. Kojima, Prog. Theor. Phys. {\bf 90}, 977 (1993).

\bibitem{McClintock:etal:2006}
J.E.~McClintock, R.~Shafee, R.~Narayan, R.A.~Remillard, S.W.~Davis, L.X.~Li, 
Astrophys.J. {\bf 652} (2006) 518-539



\bibitem{Middleton:16}
M.~Middleton,
Astrophys. Space Sci. Libr. 440 (2016) 99 
[arXiv:1507.06153] 



\bibitem{Bardeen:Horowitz:1999}
J.M.~Bardeen and G.T.~Horowitz, 
Phys. Rev. D 60 104030 
(Preprint hep-th/9905099)


\bibitem{KLR07}
H.K.~Kunduri, J.~Lucietti, and H.S.~Reall,
Class.\ Quant.\ Grav.\  {\bf 24}, 4169-4189 (2007).
 
 

\bibitem{Guica:Hartman:Song:Strominger:2009}
M.~Guica, T.~Hartman, W.~Song, A.~Strominger, 
Phys. Rev. D{\bf 80} (2009) 124008


\bibitem{Porfyriadis:Strominer:2014}
A.P.~Porfyriadis and A.~Strominger,
Phys.Rev. D{\bf 90} (2014) 044038

\bibitem{Porfyriadis:Shi:Strominger:2017}
A.P.~Porfyriadis, Y.~Shi, and A.~Strominger,
Phys.Rev. D{\bf 95} (2017) 064009

 
 
\bibitem{Figueras:Kunduri:Lucietti:Rangamani:2008}
P.~Figueras, H.K.~Kunduri, J.~Lucietti, M.~Rangamani,
Phys. Rev. D{\bf 78} (2008) 044042


\bibitem{Kunduri:Lucietti:2009}
H.K.~Kunduri and J.~Lucietti, 
J. Math. Phys. {\bf 50} (2009) 082502


\bibitem{Hollands:Ishibashi:2010}
S.~Hollands and A.~Ishibashi, 
Annales Henri Poincare {\bf 10} (2010) 1537-1557



\bibitem{Kunduri:Lucietti:2013}
H.K.~Kunduri and J.~Lucietti, 
Living Rev.Rel. 16 (2013) 8


\bibitem{Zimmerman:2017}
P.~Zimmerman, 
Phys. Rev. D{\bf 95} (2017) 124032

\bibitem{BF1}
P. Breitenlohner and D.Z. Freedman, 
Ann. Phys. 144 (1982) 249 [SPIRES].

\bibitem{BF2}
P. Breitenlohner and D.Z. Freedman, 
Phys. Lett. B {\bf 115} (1982) 197 [SPIRES].


\bibitem{Gubser2008}
S.S.~Gubser, 
Phys. Rev. D{\bf 78} (2008) 065034

\bibitem{Hartnoll:Herzog:Horowitz:2008}
S.A. Hartnoll, C.P. Herzog and G.T. Horowitz, 
JHEP 12 (2008) 015



\bibitem{Dias:Monteiro:Reall:Santos:2010}
O.J.C.~Dias, R.~Monteiro, H.S.~Reall, J.E.~Santos, 
JHEP 1011 (2010) 036


\bibitem{Aretakis:2011}
S. Aretakis, 
Commun. Math. Phys. {\bf 307}, 17 (2011)

\bibitem{Durkee:Reall:11}
M. Durkee and H. S. Reall, 
Phys. Rev. D {\bf 83}, 104044 (2011)

\bibitem{Hollands:Ishibashi:15}
S.~Hollands, A.~Ishibashi
Commun. Math. Phys. {\bf 339} (2015) 949-1002



\bibitem{Frolov:Krtous:Kubiznak:Santos:2018}
V.P.~Frolov, P.~Krtouš, D.~Kubizňák, J.E.~Santos, 
arXiv:1804.00030 [hep-th]


\bibitem{Kodama:Ishibashi:2003}
H.~Kodama, A.~Ishibashi, 
Prog. Theor. Phys. {\bf 110} (2003) 701-722


\bibitem{Ishibashi:Kodama:2003}
A.~Ishibashi, H.~Kodama, 
Prog. Theor. Phys. {\bf 110} (2003) 901-919



\end{thebibliography}
\end{document}